\documentclass[prapplied,aps,superscriptaddress,showpacs,reprint]{revtex4-2}

\usepackage{graphicx}
\usepackage{dcolumn}
\usepackage{bm}
\usepackage{SIunits}
\usepackage{color}
\usepackage[dvipsnames]{xcolor}
\usepackage[colorlinks=true, citecolor={blue!80!black}, urlcolor={blue!50!black}, linkcolor = {blue!80!black}]{hyperref}
\usepackage{appendix}
\begin{document}
	
	
\title{Electrical Properties of Selective-Area-Grown Superconductor-Semiconductor Hybrid Structures on Silicon}

\author{A. Hertel}
\affiliation{Center for Quantum Devices and Microsoft Quantum Lab--Copenhagen, Niels Bohr Institute, University of Copenhagen, 2100 Copenhagen, Denmark}
	
\author{L. O. Andersen}
\affiliation{Center for Quantum Devices and Microsoft Quantum Lab--Copenhagen, Niels Bohr Institute, University of Copenhagen, 2100 Copenhagen, Denmark}
	
\author{D. M. T. van Zanten}
\affiliation{Center for Quantum Devices and Microsoft Quantum Lab--Copenhagen, Niels Bohr Institute, University of Copenhagen, 2100 Copenhagen, Denmark}
	
\author{M. Eichinger}
\affiliation{Center for Quantum Devices and Microsoft Quantum Lab--Copenhagen, Niels Bohr Institute, University of Copenhagen, 2100 Copenhagen, Denmark}
	
\author{P. Scarlino}
\affiliation{Center for Quantum Devices and Microsoft Quantum Lab--Copenhagen, Niels Bohr Institute, University of Copenhagen, 2100 Copenhagen, Denmark}
	
\author{S. Yadav}
\affiliation{Center for Quantum Devices and Microsoft Quantum Lab--Copenhagen, Niels Bohr Institute, University of Copenhagen, 2100 Copenhagen, Denmark}
	
\author{J. Karthik}
\affiliation{Center for Quantum Devices and Microsoft Quantum Lab--Copenhagen, Niels Bohr Institute, University of Copenhagen, 2100 Copenhagen, Denmark}
	
\author{S. Gronin}
\affiliation{Department of Physics and Astronomy and Microsoft Quantum Lab--Purdue, Purdue University, West Lafayette, Indiana 47907, USA}
\affiliation{Birck Nanotechnology Center, Purdue University, West Lafayette, Indiana 47907, USA}
	
\author{G. C. Gardner}
\affiliation{Department of Physics and Astronomy and Microsoft Quantum Lab--Purdue, Purdue University, West Lafayette, Indiana 47907, USA}
\affiliation{Birck Nanotechnology Center, Purdue University, West Lafayette, Indiana 47907, USA}
	
\author{M. J. Manfra}
\affiliation{Department of Physics and Astronomy and Microsoft Quantum Lab--Purdue, Purdue University, West Lafayette, Indiana 47907, USA}
\affiliation{Birck Nanotechnology Center, Purdue University, West Lafayette, Indiana 47907, USA}
\affiliation{School of Materials Engineering, Purdue University, West Lafayette, Indiana 47907, USA}
\affiliation{School of Electrical and Computer Engineering, Purdue University, West Lafayette, Indiana 47907, USA}
	
\author{C. M. Marcus}
\affiliation{Center for Quantum Devices and Microsoft Quantum Lab--Copenhagen, Niels Bohr Institute, University of Copenhagen, 2100 Copenhagen, Denmark}
	
\author{K. D. Petersson}
\affiliation{Center for Quantum Devices and Microsoft Quantum Lab--Copenhagen, Niels Bohr Institute, University of Copenhagen, 2100 Copenhagen, Denmark}

\date{\today}
	
\begin{abstract}
				We present a superconductor-semiconductor material system that is both scalable and monolithically integrated on a silicon substrate. It uses selective area growth of Al-InAs hybrid structures on a planar III-V buffer layer, grown directly on a high resistivity silicon substrate. We characterized the electrical properties of this material system at millikelvin temperatures and observed a high average field-effect mobility of $\mu \approx 3200\,\mathrm{cm^2/Vs}$ for the InAs channel, and a hard induced superconducting gap. Josephson junctions exhibited a high interface transmission, $\mathcal{T} \approx 0.75 $, gate voltage tunable switching current with a product of critical current and normal state resistance, $I_{\mathrm{C}}R_{\mathrm{N}} \approx 83\,\mathrm{\micro V}$, and signatures of multiple Andreev reflections. These results pave the way for scalable and high coherent gate voltage tunable transmon devices and other superconductor-semiconductor hybrids fabricated directly on silicon.

\end{abstract}
	
\maketitle
	
\section{\label{sec:level1}Introduction}
	
	Recent demonstrations of superconducting qubit systems with over fifty qubits places superconducting qubits as the most advanced approach to building a quantum computer~\cite{Arute2019}. However, practically useful fault tolerant quantum computers may require several orders of magnitude more qubits all operating with high fidelity at millikelvin temperatures~\cite{Reiher2017}. This substantial challenge of scaling superconducting qubit systems motivates the exploration of novel materials platforms that might unlock new and readily extensible approaches to qubit control~\cite{Tahan2019}. In particular, recent work has demonstrated that semiconductor elements can be incorporated into superconducting systems to create superconducting qubits with gate-voltage-tunable Josephson junctions (JJ)~\cite{Larsen2015,Casparis2018,Wang2019}. This semiconductor-based JJ technology has the potential to enable superconducting qubits to naturally interface with ultra low power cryogenic CMOS control~\cite{Pauka2019}.
	
	So far, approaches to building semiconductor-based superconducting transmon qubits, known as gatemons, have either proven challenging to scale in the case of VLS-nanowire-based qubits~\cite{Larsen2015} or had coherence times that were limited by losses in the underlaying III-V substrate in the case of two-dimensional-electron-gas-based gatemons~\cite{Casparis2018}. In this work, we present a novel materials system that brings together the salient features of these earlier approaches by using deterministic selective-area-growth of Al-InAs hybrid structures~\cite{Krizek2018} on a low loss Si substrate. We study the electrical properties of this hybrid system with respect to the requirements for gatemon qubits. In particular, disorder both within the semiconductor JJ channel and at the superconductor-semiconductor interface can lead to subgap states that can act as an additional decoherence channel for the qubit~\cite{Chang2015, Bilmes2017}. We focus on the hybrid Al-InAs system as the basis for our semiconductor JJs. InAs supports high mobility transport and the epitaxial combination of Al and InAs has been shown to form a high quality, high transparency superconductor-semiconductor interface~\cite{Chang2015} along with the highest coherence gatemon devices~\cite{Casparis2016, Luthi2018}. We integrate these materials on a Si platform that has been shown to support high coherence superconducting qubits~\cite{Dunsworth2017, Burnett2019} and offers compatibility with silicon CMOS technology.

	In order to characterize our system, first, we extract mobilities for the semiconductor channel using field effect transistor devices (FETs). Next, we probe the superconductor-semiconductor interface quality through transport spectroscopy of normal-conductor-insulator-superconductor (NIS) junctions. Finally, we characterize the quality of JJs in our material system by extracting the $I_{\mathrm{C}}R_{\mathrm{N}}$ product (where $I_{\mathrm{C}}$ and  $R_{\mathrm{N}}$ are the JJ critical current and normal state resistance) and the junction transparency.

	\begin{figure}[t]
		\includegraphics[width=1\columnwidth]{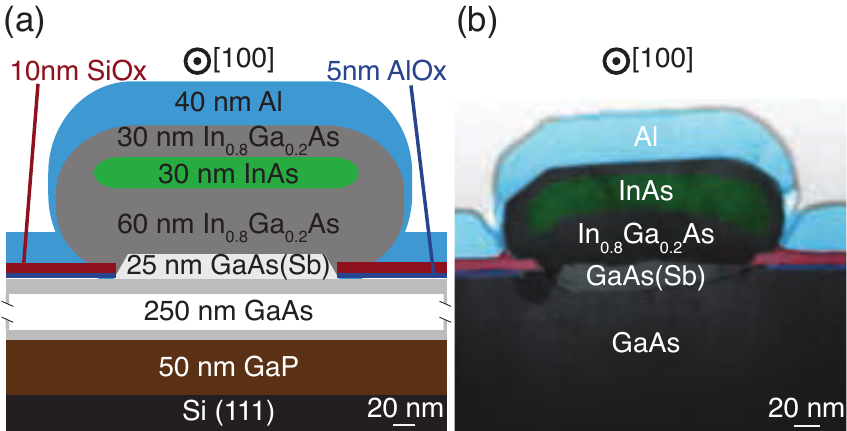}
		\caption{\label{fig:Fig1} \textbf{(a)} Schematic of the material stack. The InAs quantum well (green) is proximitized by the superconducting Al (blue). (\textbf{b)} False-colored scanning transmission electron micrograph of the material stack.}
		\label{fig:material_fig}
	\end{figure}

\section{\label{sec:level1b}Material Stack and Devices}

	Figure \ref{fig:material_fig} shows the material stack that is studied in this work. It was grown on a $4\degree$ miscut (111)-Si substrate with resistivity $\rho > 1\,\mathrm{k\Omega\cdot cm}$. To grow the high-quality InAs layer, GaAs/GaP/Si substrates consisting of buffer layers of $\sim 50\,\text{nm}$ of GaP and $\sim 250\,\text{nm}$ of GaAs were used. These buffers were deposited using metal organic chemical vapor deposition (MOCVD) techniques and are commercially available~\cite{Nemeth2008}. The GaP-on-Si buffer technique enables anti-phase defect free growth of III-V materials on non-polar Si substrate \cite{Doescher2008}. The GaAs buffer serves multiple purposes. First, it reduces the lattice mismatch between the InAs and Si (from ~11.4\% to 7.2\%). Second, it provides a smooth and high-surface quality virtual GaAs substrate on which high-quality in-plane InAs nanowires can be grown using selective-area growth techniques (SAG) \cite{Krizek2018}. Third, owing to its large bandgap, it provides a good electrical isolation between the InAs nanowires and the Si substrate. The total thickness of the GaP/GaAs buffer stack was limited to $\sim 300\,\text{nm}$ to simplify integration of our superconductor-semiconductor material system with any low loss qubit circuit components that are fabricated directly on the underlying Si substrate. 
	
	We used selective area growth techniques~\cite{Krizek2018} to grow InAs structures shaped as planar nanowires on the GaAs surface. Thin films of AlOx and SiOx were deposited on the wafer using atomic layer deposition and plasma-enhanced chemical vapor deposition, respectively. Openings in these dielectric layers were then defined by standard electron beam lithography and selective etching of the SiOx and AlOx using reactive-ion etching and a wet-etch solution, respectively. The subsequent growth steps took place by molecular beam epitaxy (MBE) in ultra high vacuum prepared with high purity processes \cite{Gardner2016} and the semiconductor heterostructures were then grown selectively in the dielectric openings where the GaAs surface was exposed. The SAG heterostructure was designed to increase the interface quality between the GaAs and InAs layers in order to improve the overall InAs material quality. First, an Sb-dilute GaAs buffer layer with flat top facets was grown. This layer enables significant elastic strain relaxation of the InAs~\cite{Krizek2018}. Then a layer of $\text{In}_{0.8}\text{Ga}_{0.2}\text{As}$ was grown to bridge the lattice mismatch to the InAs. To help prevent surface damage from subsequent device processing steps impacting transport characteristics, a top barrier layer of $\text{In}_{0.8}\text{Ga}_{0.2}\text{As}$ was grown on the InAs. Finally, a blanket Al layer was deposited \textit{in situ} \cite{Krogstrup2015} to ensure a high-quality interface between the Al and the semiconductor heterostructure. A false-colored scanning transmission electron microscope image of the top part of the material stack is shown in Fig.~\ref{fig:material_fig}b. 
	
\begin{figure}[t]
	\includegraphics[width=1\columnwidth]{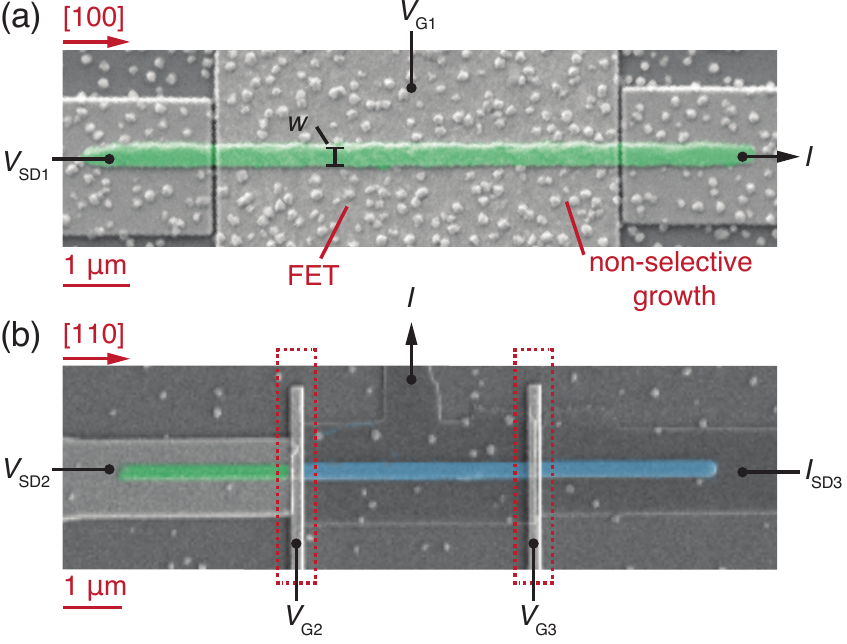}
	\caption{\label{fig:Fig1} False-colored scanning electron micrographs of the device types measured in this work. False colors indicate nanowire segments with Al removed (green) and segments covered with Al (blue). \textbf{(a)} Field-effect transistor (FET) device. \textbf{(b)} Combined NIS spectroscopy and SSmS Josephson junction device. The middle segment serves as ground for both NIS spectroscopy and junction measurements. On some wafers, poor selectivity during SAG leads to material deposition on the oxide mask \cite{Aseev2019} which is visible as small grains in both SEMs.}
	\label{fig:devices}
\end{figure}

	We fabricated three different types of devices to study the properties of our material system: FETs, NIS junctions and superconductor-semiconductor-superconductor Josephson junctions (SSmS JJs) (Fig.~\ref{fig:devices}). NIS and SSmS JJ devices are used to study the superconductor-semiconductor hybrid system.~They were fabricated in pairs using a single nanowire such that both devices shared a common superconducting segment (Fig.~\ref{fig:devices}b). This was achieved by selectively etching the Al layer and depositing a normal-metal contact. The semiconducting segments were tuned by topgates, separated from the nanowire by gate dielectric.
	
	FET devices are used to study the material quality of the conducting InAs layer exclusively. To fabricate these devices we used standard electron beam lithography. In a first step Al was removed from the nanowires by selective wet etching. Subsequently, contacts, gate dielectric, and a topgate were deposited (see Appendix A for additional device fabrication details). The length of the gated channel for all devices shown in this work is $L_{\mathrm{FET}} = 6\,\mathrm{\micro m}$. All measurements in this work were performed at $30\,\text{mK}$ unless stated otherwise.  
		
\section{\label{sec:level2}Field-effect measurements}
	
\begin{figure}[b!]
	\includegraphics[width=0.9\columnwidth]{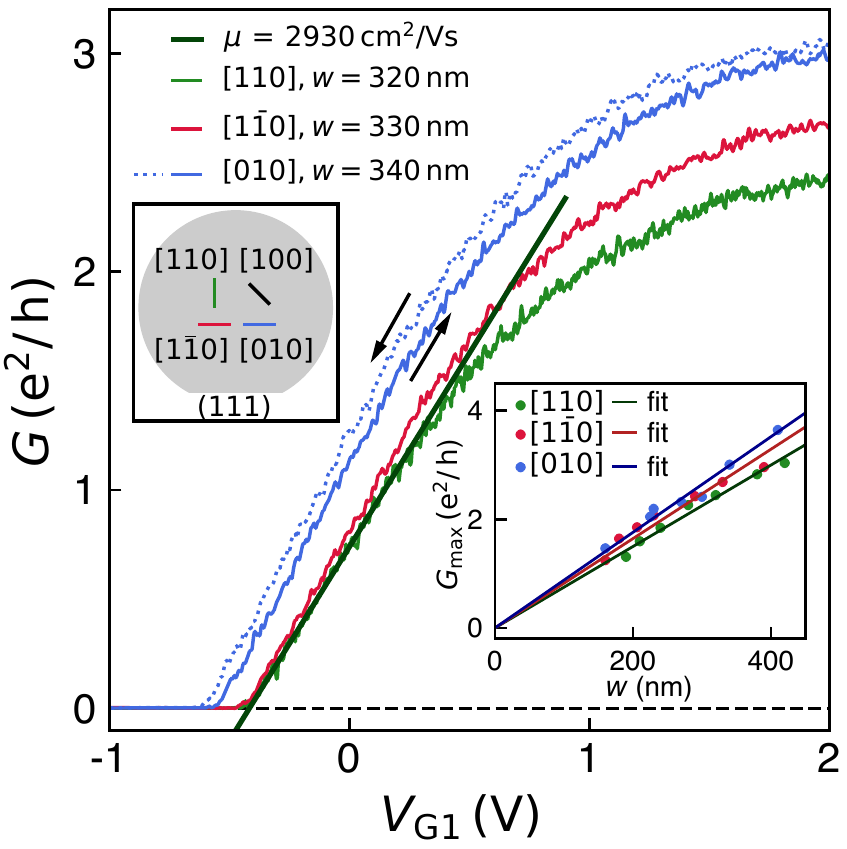}
	\caption{\label{fig:Fig2} Differential conductance $G$ as a function of the gate voltage $V_{\mathrm{G1}}$ of an up- and down sweep of a single nanowire (blue) and up sweeps of two nanowires with a different orientation and similar width $w$. A fit to the linear region of the green pinch-off curve is shown and used to determine the field-effect mobility according to Eq. 1. \textbf{Upper left inset}: schematic showing orientation of nanowires relative to the major flat of the Si-substrate. \textbf{Lower right inset}: conductance in the fully open regime as a function of $w$. Solid lines indicate fits of the form $G_\mathrm{max} = g\prime \cdot w$, where $g\prime$ is the conductance per unit width of the nanowires in the fully opened regime. All the data shown in this figure have been corrected for a line resistance $R_{\mathrm{line}} = 4.9\,\mathrm{k\Omega}$.}
	\label{fig:FET}
\end{figure}
	
	A typical figure of merit for the quality of a semiconductor is the field-effect mobility as it is limited by defects in the semiconductor, gate dielectric layers and interfaces~\cite{Guenel2014}. The field-effect mobility can be extracted from the transconductance $\mathrm{d}G/\mathrm{d}V_{\mathrm{G1}}$ which we obtained from measurements of the differential conductance $G$ as a function of gate voltage $V_{\mathrm{G1}}$ for a total of 24 FET devices (similar to the one represented in Fig.~\ref{fig:devices}a) using standard lock-in techniques. The measured nanowires were grown along the $[100], [110]$ and $[1\bar{1}0]$ directions of the underlying Si substrate (Fig.~\ref{fig:FET} inset). 
	
	Figure \ref{fig:FET} shows typical differential conductance measurements sweeping $V_{\mathrm{G1}}$ in both positive and negative directions. Nanowires for all different orientations could be fully pinched-off and showed a small hysteresis of about ~$\sim 0.05\,V$, indicating high-quality interfaces between the electrically active InAs channel and adjacent layers~\cite{Krizek2018}. As depicted in Fig.~\ref{fig:FET},  the mean field effect mobility $\mu$ is extracted from a fit to the linear region of the conductance with the highest slope~\cite{Guel2015}. It can be estimated by
	\begin{equation}
	\frac{\mathrm{d}G}{\mathrm{d}V_{\mathrm{G1}}}\Bigg|_{\text{max}} = \frac{\mu C_{\mathrm{G}}}{L_{\mathrm{FET}}^2},
	\end{equation}
	where $L_{\mathrm{FET}} = 6\,\mathrm{{\micro m}}$  is the length of the gated channel and $C_{\mathrm{G}}$ is the gate capacitance that is estimated from finite-element simulations (see Supplemental Material, Section I). The extracted averaged value for all devices is $\mu_{\mathrm{avg}} =  (3200 \pm 300)\,\mathrm{cm^2/Vs}$. Here, the uncertainty of the simulated capacitance is neglected and the given error is the statistical error of all measured devices, where up and down sweeps of the same device were considered to be two independent measurements. The value $\mu_{\mathrm{avg}}$ is comparable but lower than values reported for InAs SAG nanowires grown directly on GaAs~\cite{Krizek2018} and InAs VLS-nanowires that are grown strain-relaxation-free~\cite{Schroer2010, Wang2013}. Previous work has shown that the low-temperature field effect mobility of undoped III-V nanowires is typically limited by crystal defects~\cite{Schroer2010, Gupta2013, Sourribes2014} or surface effects~\cite{Wang2013, Gupta2013}. Further work would be needed to understand the dominant electron scattering mechanism in our material stack in order to further optimize the field-effect mobility. Based on the threshold voltages $V_{\mathrm{th}}$ and the volume of the conducting channel $v_{\mathrm{ch}}$ (Fig.~\ref{fig:material_fig}) we estimate a mean carrier density at zero gate voltage using $n = C_{\mathrm{G}}V_{\mathrm{th}}/(\mathrm{e}v_{\mathrm{ch}})$. For the different nanowire orientations, we estimate $n_{[0\bar{1}0]} = 3.0\cdot 10^{17}\,\mathrm{cm^{-3}}$, $n_{[100]} = 3.4\cdot 10^{17}\,\mathrm{cm^{-3}}$, and $n_{[010]} = 4.8\cdot 10^{17}\,\mathrm{cm^{-3}}$. Nanowires along the $[010]$ direction exhibit the highest charge carrier density (lowest $V_{\mathrm{th}}$) and the highest conductance when the conducting channel is fully opened (for gate voltages $V_{\mathrm{G1}}>2\,{\mathrm{V}})$ (Fig.~\ref{fig:FET}, lower inset).

\section{\label{sec:NIS} Induced superconducting gap}
	
\begin{figure}[t!]
	\includegraphics[width=1\columnwidth]{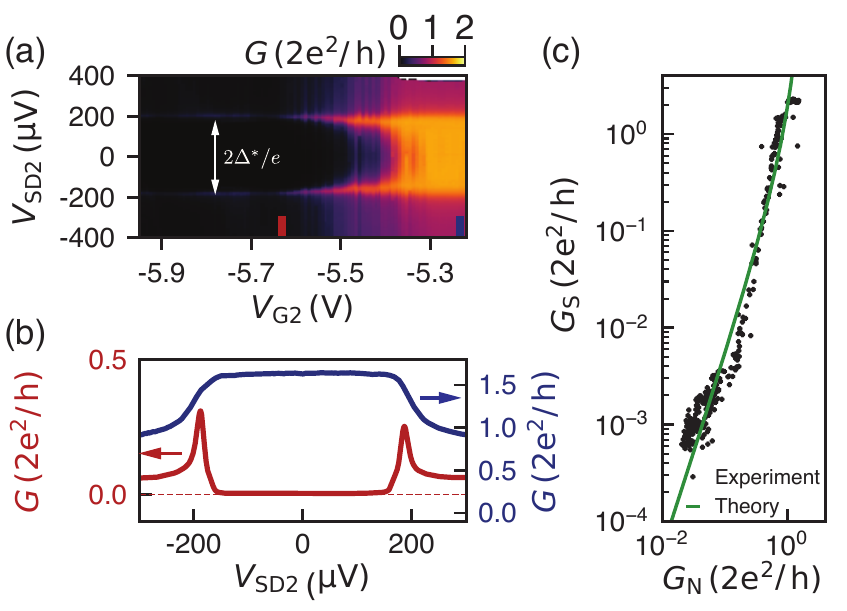}
	\caption{\label{fig:wide} \textbf{(a)} Differential conductance $\mathrm{d}I/\mathrm{d}V$ of nanowire A as a function of gate voltage $V_\mathrm{G2}$ and source-drain voltage $V_{\mathrm{sd}}$. \textbf{(b)} Vertical cuts in the tunneling regime (red) and open regime (blue) at the positions indicated by the colored rectangles in (a). The induced superconducting gap $\Delta^{*} \approx 190\,\mathrm{{\micro eV}}$ is estimated from the peak-to-peak distance in the tunneling regime. \textbf{{(c)}} Averaged differential conductance at zero source-drain voltage $G_{\mathrm{S}}$ versus averaged differential conductance at finite source-drain voltage $G_{\mathrm{N}}$ ($V_{\mathrm{SD2}}) = -0.53\,\mathrm{mV}$. The green line is the theoretically predicted conductance in an Andreev enhanced QPC with no fitting parameters (Eq.(2)).}
	\label{fig:NIS}
\end{figure}

	To study the interface quality between the superconducting Al and InAs and the induced superconducting gap $\Delta^{*}$ in the InAs, we used the NIS device introduced previously in Fig.~\ref{fig:devices}a and fabricated on nanowire A. The device was measured with standard lock-in techniques with unused contacts left floating so the third segment on the nanowire did not affect the measurement. As shown in Fig.~\ref{fig:NIS}a, we depleted the bare InAs segment by applying a negative gate voltage $V_{\mathrm{G2}}$ to create a tunnel barrier and measured the differential conductance $G$ of the device as a function of voltage bias $V_{\mathrm{SD2}}$. 
	
   Figure~\ref{fig:NIS}b (red curve) shows a measurement in the tunneling regime. The conductance $G$ is strongly suppressed between two symmetric peaks. The peak positions are independent of the gate voltage and indicate an induced gap $\Delta^{*} \approx 190\,\mathrm{\micro eV}$ (see Fig.~\ref{fig:NIS}a). At higher $V_{\mathrm{G2}}$ (more open barrier) the subgap conductance $G_{\mathrm{S}}$ (measured at $V_{\mathrm{SD2}}=0 \,\mathrm{V}$) is increased compared to the above-gap conductance $G_{\mathrm{N}}$ for $\left|V_{\mathrm{SD2}}\right| \gtrsim200 \,\mathrm{\micro eV}$ (Figure ~\ref{fig:NIS}b, blue curve). 
   Both observations, suppressed ($G_{\mathrm{S}}\ll G_{\mathrm{N}}$) and enhanced zero-bias-conductance ($G_{\mathrm{S}}> G_{\mathrm{N}}$), can be explained in the framework of the Blonder-Tinkham-Klapwijk (BTK) formalism~\cite{Blonder1982}. This theory describes the charge transfer through an NS interface by Andreev reflection using a single parameter, the dimensionless barrier parameter $Z$. The limit $Z=0$ corresponds to a perfect interface where $G_{\mathrm{S}} = 2G_{\mathrm{N}}$ is expected as every charge carrier is Andreev reflected at the interface. The limit $Z \rightarrow \infty$ corresponds to a perfect tunnel barrier where the conductance is directly proportional to the density of states in the proximitized region. Thus, changing $V_{\mathrm{G2}}$ in the experiment corresponds to modifying the $Z$-parameter. To further study the transport across the NS interface we compare the experiment to
	\begin{equation}
	G_{\mathrm{S}} = 2 G_0 \frac{G_{\mathrm{N}}^2}{(2G_0 - G_{\mathrm{N}})^2},
	\end{equation}
	the theoretical prediction for a quantum point contact (QPC) with a single channel that correlates $G_{\mathrm{S}}$ to $G_{\mathrm{N}}$~\cite{Beenaker1992} without any free parameters and $G_0 = 2e^2/h$. The measurement was repeated on the same nanowire using a DC setup to measure small differential conductance values ($G_{\mathrm{S}} < 10^{-2}\,\mathrm{e^2/h}$). Here, the differential resistance is obtained from calculating the numerical derivative of the DC data. The result is shown in the Supplemental Material (Fig.~S4) and used to construct Fig.~\ref{fig:NIS}c. The experimental data follows the theoretical prediction (green line in Fig.~\ref{fig:NIS}c). We therefore conclude the presence of an induced hard superconducting gap in the InAs. The small deviation could be the manifestation of a non-zero normal scattering probability or the presence of a multiple of conducting channels with transmission probability below 1. The presence of multiple conducting channels is evident in the plateau region (Supplemental Material Fig.~S4d)  with enhanced zero-bias conductance, $G_{\mathrm{N}}$ ($V_{\mathrm{G2}}>-5.3\,\mathrm{V}$ in Fig.~\ref{fig:NIS}a), that is not quantized at $2\mathrm{e^2/h}$, contrary to the expectations for a single perfectly transmitting channel.

\section{\label{sec:critical_current}Nanowire Josephson junctions}
	
\begin{figure}[t!]
	\includegraphics[width=1\columnwidth]{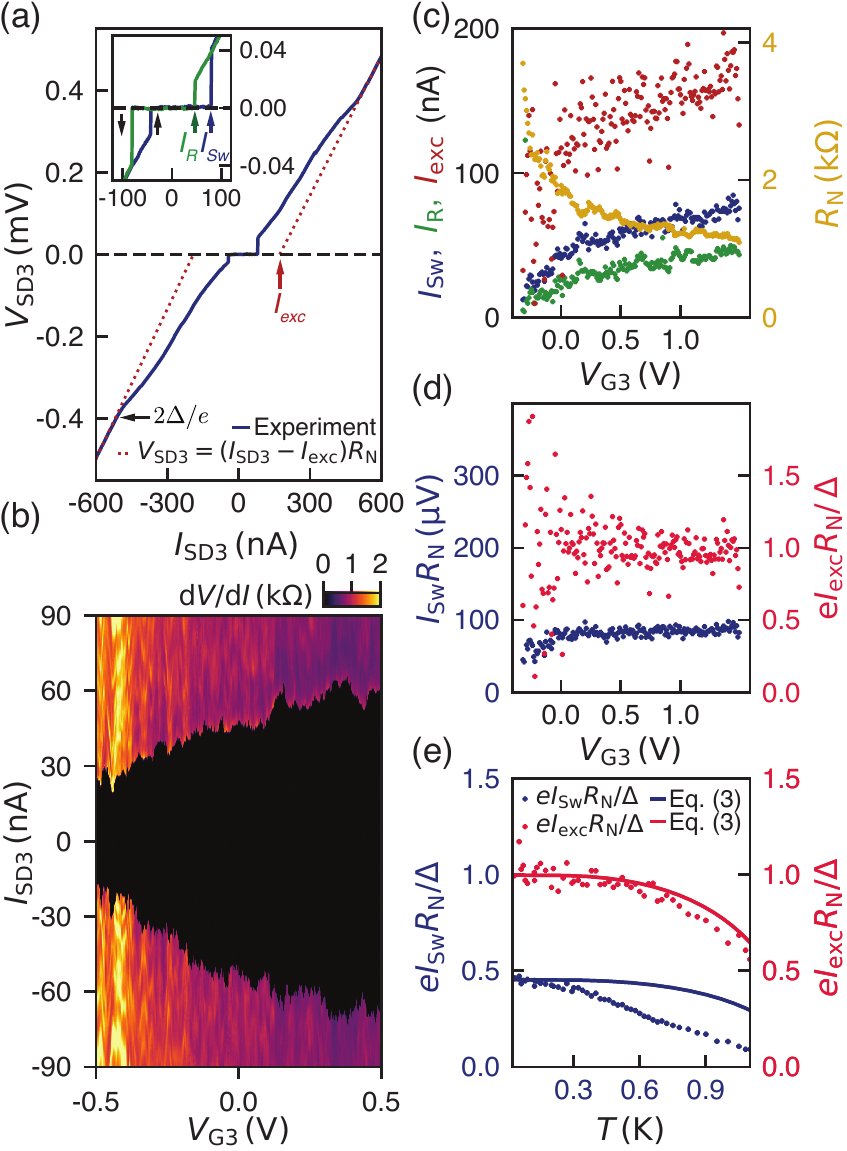}
	\caption{\label{fig:Fig4}\textbf{(a)} Measured voltage drop across the Josephson junction $V_{\mathrm{SD3}}$ as a function of applied current $I_{\mathrm{SD3}}$ for $V_{\mathrm{G3}} = 1.5\,\mathrm{V}$ for nanowire C. The normal state resistance $R_{\mathrm{N}}$ and excess current $I_{\mathrm{exc}}$ are extracted by fitting to the normal state for voltages $V_{\mathrm{SD3}} > 2\Delta/e$. From these measurements we estimate $\Delta \approx 200\,\mathrm{\micro eV}$ at $T=20\,\mathrm{mK}$. The inset shows the region around the superconducting plateau with an up- and down sweep of the current bias. The position  of the switching current $I_{\mathrm{Sw}}$ and retrapping current $I_{\mathrm{R}}$ are indicated by arrows. \textbf{(b)} Differential resistance $\mathrm{d}V/\mathrm{d}I$ as a function of applied current $I_{\mathrm{SD3}}$ and gate voltage $V_{\mathrm{G3}}$ at $T=20\,\mathrm{mK}$ for nanowire C.  \textbf{(c)} Extracted values for $R_{\mathrm{N}}$, $I_{\mathrm{exc}}$, $I_{\mathrm{Sw}}$ and $I_{\mathrm{R}}$ extracted for nanowire B from the dataset shown in the Supplemental Material. \textbf{(d)} The $I_{\mathrm{sw}}R_{\mathrm{N}}$ product and excess current $I_{\mathrm{exc}}$ multiplied by the ratio $\mathrm{e}R_{\mathrm{N}}/\Delta$ for $\Delta = 200\,\mathrm{\micro eV}$ extracted from \textbf{(c)}.  \textbf{(e)} $\mathrm{e}I_{\mathrm{exc}}R_{\mathrm{N}}/\Delta$ and $\mathrm{e}I_{\mathrm{Sw}}R_{\mathrm{N}}/\Delta$ as a function of temperature $T$ calculated from data shown in the Supplemental Material [see Fig.~S5]. Solid lines indicate the temperature dependence expected from BCS interpolation (Eq. 3).} 
	\label{fig:SIS}
\end{figure}

	We use the SSmS JJ device (Fig.~\ref{fig:devices}b) to study the Josephson junction formed in the material system. We characterize the junction extracting several junction parameters, the switching current, $I_{\mathrm{Sw}}$, the retrapping current, $I_{\mathrm{R}}$, the excess current $I_{\mathrm{exc}}$, the normal state resistance $R_{\mathrm{N}}$ and the superconducting gap $\Delta$ as a function of gate voltage $V_{\mathrm{G3}}$ and temperature $T$. All parameters are extracted as shown in Fig.~\ref{fig:SIS}a for a typical IV-curve. The junction switches to a dissipative state at switching current $I_{\mathrm{Sw}}$ as the bias current, $I_{\mathrm{SD3}}$, is swept from zero. Sweeping $I_{\mathrm{SD3}}$ back towards zero, the junction switches back to the superconducting state at the retrapping current, $I_{\mathrm{R}}$, visible as hysteretic behavior (Fig.~\ref{fig:SIS}a, inset). The normal state resistance, $R_{\mathrm{N}}$, and excess current, $I_{\mathrm{exc}}$, are extracted at high current bias where $V_{\mathrm{SD3}} > 2\Delta/e$, and the junction is driven into the normal conducting state. We estimate the superconducting gap, $\Delta \approx 200\,{\mathrm{\micro eV}}$, from the visible transition to the normal conducting state. This value is similar to the estimated induced superconducting gap of the SIN device, $\Delta^{*} \approx 190\,{\mathrm{\micro eV}}$. The critical current is gate tunable as demonstrated by Fig.~\ref{fig:SIS}b, where the differential resistance $\mathrm{d}V/\mathrm{d}I$ is shown as a function of applied current $I_{\mathrm{SD3}}$ and gate voltage $V_{\mathrm{G3}}$. Furthermore, IV curves exhibit subgap features in the resistive state for $V_{\mathrm{SD3}} < 2\Delta/e$ that result from multiple Andreev reflections (MARs)~\cite{Octavio1983,Goloubov1996}, discussed in the next section. Figure~\ref{fig:SIS}c shows the extracted junction parameters for nanowire B over a wide gate range from $V_{\mathrm{G3}} = -0.3\,\mathrm{V}$, where the nanowire junction is almost closed ($I_{\mathrm{Sw}} \approx 0\,\mathrm{nA}$), to $V_{\mathrm{G3}} = 1.5\,\mathrm{V}$, where the junction is fully opened. While $I_{\mathrm{Sw}}/I_{\mathrm{R}} \approx 1.0$ in the closed regime, we find ratios $I_{\mathrm{Sw}}/I_{\mathrm{R}}\approx 2.0$ for gate voltages around $V_{\mathrm{G3}} = 1.5\,{\mathrm{V}}$, indicating that the nanowire junction may be underdamped \cite{Tinkham2004}. 
	
	To further characterize the JJ and its superconductor-semiconductor interface quality, we extract the product $I_{\mathrm{Sw}}R_{\mathrm{N}}$ and the normalized excess current $eI_{\mathrm{exc}}R_{\mathrm{N}}/\Delta$ \cite{Octavio1983} (Fig.~\ref{fig:SIS}d). Here, we use $I_{\mathrm{Sw}}$ as a proxy for the critical current, $I_{\mathrm{C}}$, noting that the measured switching current can be smaller than the actual critical current of the junction due to premature switching~\cite{Tinkham2004} or coupling to the electromagnetic environment~\cite{Jarillo-Herrero2006}. The normalized excess current $eI_{\mathrm{exc}}R_{\mathrm{N}}/\Delta$ is used to estimate the BTK barrier parameter $Z$ for NS interface scattering~\cite{Blonder1982} and is connected to the interface transmission $\mathcal{T} = (1+Z^2)^{-1}$. Hence, a high value of $\mathcal{T}$ is an indication of a high superconductor-semiconductor interface quality. For $V_{\mathrm{G3}} > 0.1\,\mathrm{V}$ we extract an averaged excess current $eI_{\mathrm{exc}}R_{\mathrm{N}}/\Delta =  0.99\pm 0.15$ and estimate $Z = 0.58 \pm 0.06$~\cite{Flensberg1988}, leading to $\mathcal{T} = 0.75 \pm 0.04$. A high transparency is consistent with the enhanced zero bias conductance observed with tunneling spectroscopy and with the relatively high value of the induced superconducting gap in the InAs compared to the Al gap~\cite{Blonder1982, Octavio1983}. For $V_{\mathrm{G3}} < 0\,\mathrm{V}$, $I_{\mathrm{Sw}}R_{\mathrm{N}}$ is reduced and the spread in $eI_{\mathrm{exc}}R_{\mathrm{N}}/\Delta$ increases. For $V_{\mathrm{G3}} < -0.35\,\mathrm{V}$ negative and positive values for $I_{\mathrm{exc}}$ are extracted. Similar observations have previously been explained by quantum dots forming in the junction, which makes the transport dominated by an interplay between superconductivity and Coulomb interactions~\cite{Eiles1994,Ridderbos2018}. The extracted average value $I_{\mathrm{Sw}}R_{\mathrm{N}} = (83.3\pm\, 5.8)\mathrm{\micro V}$ is significantly smaller than the theoretical value for a short diffusive junction $I_{\mathrm{C}}R_{\mathrm{N}} = 1.32\pi\Delta/2e\approx415\,\mathrm{\micro V}$~\cite{Tafuri2019} but in good agreement with values previously measured in Al-InAs-VLS nanowires~\cite{Abay2014,Guenel2014,Doh2005,Montemurro2015} and other superconductor-semiconductor hybrid systems~\cite{Frielinghaus2010,Ojeda-Aristizabal2009,Jarillo-Herrero2006}. Similar to previous studies, the origin of a low $I_{\mathrm{Sw}}R_{\mathrm{N}}$ product in these structures is not well understood due to an insufficient understanding of electrodynamics and dissipation mechanisms in these junctions~\cite{Tafuri2019}. 

	Next we focus on the temperature dependence of junction parameters. Figure~\ref{fig:SIS}e shows
	$eI_{\mathrm{Sw}}R_{\mathrm{N}}$ and  $eI_{\mathrm{exc}}R_{\mathrm{N}}$, both normalized by the gap $\Delta$ extracted at base temperature, as a function of temperature for nanowire B at a fixed gate voltage $V_{\mathrm{G3}} = 1.5\,\mathrm{V}$.  The temperature dependence of the superconducting gap $\Delta(T)$ according to the BCS theory is also plotted against the experimental data, using the interpolation formula~\cite{Enss2005}:
	\begin{equation}
	\Delta(T) = \Delta \tanh\left(1.74\sqrt{\frac{\Delta}{1.76k_{\mathrm{B}}T}-1}\right),
	\end{equation}
	where we take $\Delta = 200\,\mathrm{\micro eV}$ extracted at $T = 20\,\mathrm{mK}$. The good agreement for the excess current over the entire temperature range suggests that the excess current is dominated by Andreev reflections at the SN interface, as has previously been observed for VLS InAs nanowires~\cite{Abay2014}. Similarly, we would expect the temperature dependence of $I_{\mathrm{Sw}}R_{\mathrm{N}}$ to follow the temperature dependence of the superconducting gap given that we estimate the junction to be in the short diffusive limit, $l_e \ll L \ll~\xi_{\mathrm{diff}}$, where $l_e$ is mean free electron path, $L$ is the junction length, and $\xi_{\mathrm{diff}}$ is the superconductor coherence length, see Appendix B~\cite{Tafuri2019}. The stronger reduction with temperature is qualitatively consistent with predictions for long diffusive SSmS junctions $(\xi_{\mathrm{diff}} \ll L)$~\cite{Kupriyanov1999, Dubos2001}, although similar behavior has been reported before for InAs-based JJs with comparable lengths, critical currents, and mean free paths~\cite{Abay2014, Mayer2019} .
	
\section{\label{sec:MAR}Multiple Andreev reflections}

\begin{figure}[t!]
	\includegraphics[width=1\columnwidth]{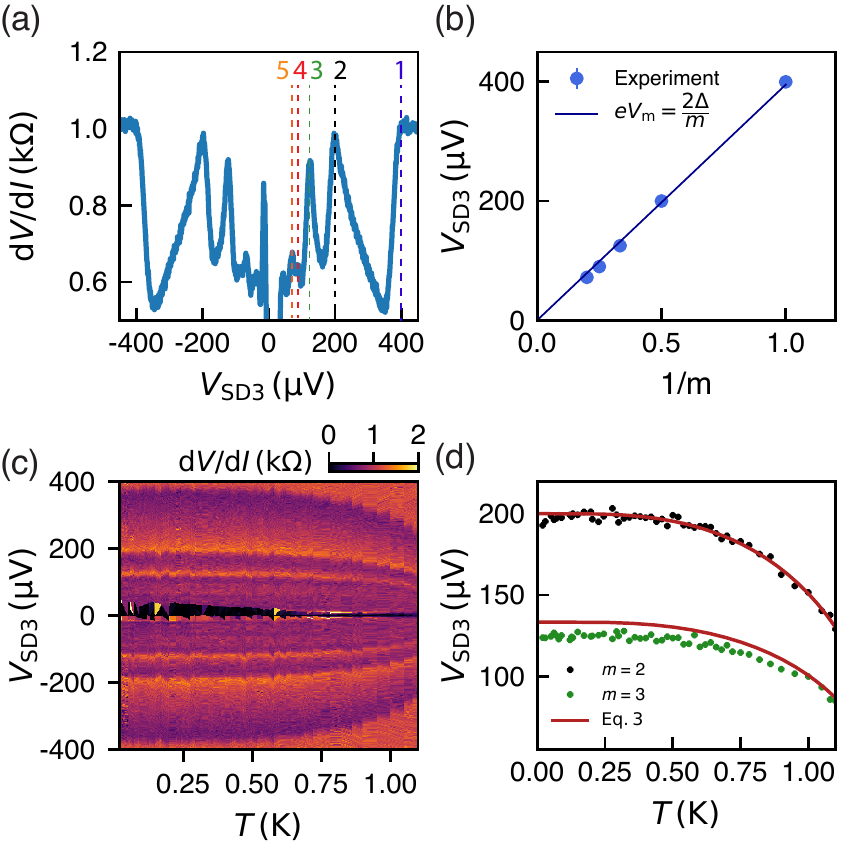}
	\caption{\label{fig:Fig5}\ \textbf{(a)} Averaged differential resistance $\mathrm{d}V/\mathrm{d}I$ for $1.3\,\text{V} < V_{\mathrm{G2}} < 1.5\,\text{V}$. Visible MAR peak positions for orders $m=1-5$ are indicated by vertical dotted lines for nanowire C. \textbf{(b)} Fit to the MAR peak position yielding a superconducting gap $\Delta \approx 200\,\mathrm{\micro eV}$. \textbf{(c)} $\mathrm{d}V/\mathrm{d}I$ as a function of measured voltage drop across the Josephson junction $V_{\mathrm{sd}}$ and sample temperature $T$. \textbf{(d)} MAR peaks for $m=2,3$ as a function of temperature. Red lines indicate the expected MAR peak positions based on $\Delta(T)$ (Eq. 3) with $\Delta = 200\,\mathrm{\micro eV}$.} 
	\label{fig:MAR}
\end{figure}

	 Subgap features in IV curves of SSmS junctions are often considered to be a manifestation of multiple Andreev reflection (MAR) processes and an indication of high overall junction transmission. Figure~\ref{fig:MAR}a shows the differential resistance averaged over a gate range $1.3\,\text{V} < V_{\mathrm{G3}} < 1.5\,\text{V}$ as a function of the voltage across the junction, $V_{\mathrm{SD3}}$, for nanowire C. For highly transparent semiconducting junctions with a few conducting channels, MAR features are expected to be visible as peaks in the differential resistance  at voltage drops $eV_{\mathrm{m}} = 2\Delta/m$~\cite{Averin1995, Kjaergaard2017, Hendrickx2019}, where $m$ denotes the MAR order. The vertical lines in Fig.~\ref{fig:MAR}a indicate the peak positions for $\Delta = 200\,\mathrm{\micro eV}$ and their corresponding MAR order $m$. From a fit to the extracted MAR positions, we obtain a superconducting gap $\Delta = (198\pm 3)\mathrm{\micro eV}$ (Fig.~\ref{fig:MAR}b). This value is in good agreement with the gap we estimate from the transition to the normal conducting state (Fig.~\ref{fig:SIS}a). 
	 
	We plot the temperature dependence of the differential resistance in Fig.~\ref{fig:MAR}c to confirm the superconducting nature of the MAR features. The MAR peak positions are extracted and separately plotted in Fig.~\ref{fig:MAR}d for $m = 2,3$. The red lines show the predicted MAR position for a BCS like gap (Eq. 3). The experimental data shows good agreement with BCS theory for $m = 2$. In the case $m=3$, the data lie systematically below the predicted position over the entire temperature range. Similar results have been obtained in other superconductor-semiconductor structures~\cite{Kjaergaard2017,Ridderbos2019}. The experimental data can potentially be better described by modeling the temperature dependence of the induced superconducting gap quantitatively as in Ref.~\cite{Kjaergaard2017}.\\
	
\section{\label{sec:conclusion}Discussion and Conclusions}
	We have characterized the different electrical properties of a novel superconductor-semiconductor materials system that uses selective-area-grown InAs with epitaxial Al on a silicon substrate. We find field effect mobilities for the InAs channel are lower but comparable with VLS-nanowires, with the crucial advantage that the selective-area-grown structures are readily scalable. We find a high quality interface between the Al and InAs as evidenced by an induced hard superconducting gap, high transparency of Josephson junctions and signatures of multiple Andreev reflections. Josephson junction exhibit a gate tunable switching current with an $I_{\mathrm{C}}R_{\mathrm{N}}$ product lower than the superconducting gap of the Al but comparable to Josephson junctions fabricated from other superconductor-semiconductor structures. The reduced $I_{\mathrm{C}}R_{\mathrm{N}}$-product in this system and similar systems is currently not well understood. 
	
	Our material system is a promising platform for scalable and high coherence gatemon devices. The hard superconducting gap and high junction transparency indicate the absence of disorder-related subgap states that can cause decoherence. The gate-tunable critical current of roughly $50\,\mathrm{nA}$ is sufficiently high to make gatemons with qubit frequencies up to around $6-8\,\mathrm{GHz}$ for typical charging energies $E_{\mathrm{C}}/h \sim 200 - 300\,\mathrm{MHz}$. The gatemon JJs could be fabricated on small mesa structures of the Al/III-V stack while low loss qubit capacitors and other readout and control components could be fabricated directly on the high resistivity silicon substrate. Moreover, with further improvement, this materials systems may also be suitable for other hybrid qubits, including protected superconducting qubits~\cite{Larsen2020} and topological qubits~\cite{Lutchyn2018}.
	
\begin{acknowledgments}
		We thank E. Yucelen for scanning transmission electron microscope sample preparation and imaging. This work was supported by Microsoft Corporation, the Danish National Research Foundation, and the Villum Foundation.
\end{acknowledgments}
	
\begin{appendix}
\section{Material and device fabrication}
	All three devices were fabricated on the same chip with standard electron beam lithography techniques. In a first step, Josephson junctions were defined by selectively wet etching $L \sim 120\,\mathrm{nm}$ long segments of the $\sim 40\,\mathrm{nm}$ thick Al film on the nanowires (Fig.~\ref{fig:devices}b). In a second step, Al was globally removed from the nanowires that were used for FET devices. Then, contacts were defined in a lift-off process. The chip was placed in an evaporation chamber and a 30s argon ion mill was performed in situ to ensure a low contact resistance followed by the evaporation of Ti/Au (5nm, 150nm). Next, $15\,\text{nm}$ of HfOx was deposited globally using atomic layer deposition as gate dielectric. For the final step Ti/Au (5nm, 150nm) was evaporated to form topgates.

\section{Coherence Length Estimate}
	
	We estimate the coherence length based on the average charge carrier density, $n$, and field-effect mobility, $\mu$, values estimated from the FET measurements (see Section~\ref{sec:level2}), considering only nanowires with the same orientation and width as nanowire B. First, we estimate the Fermi velocity, $v_{\mathrm{F}} = \hbar v_{\mathrm{F}}/m^{\ast}$, where we take the three-dimensional expression for the Fermi wavenumber, $k_{\mathrm{F}}=(6n\pi^2)^{1/3}$, and use the bulk value for the effective electron mass in InAs $m^{\ast} = 0.026m_{\mathrm{e}}$ (where $m_{\mathrm{e}}$ is the free electron mass). This gives a Fermi velocity $v_{\mathrm{F}} = 1.23\cdot 10^6\,\mathrm{m/s}$, close to the bulk InAs value $v_{\mathrm{F,bulk}} = 1.3\cdot 10^6\,\mathrm{m/s}$. We estimate a mean free electron path $l_{\mathrm{e}} = (\mu m^{\ast} v_{\mathrm{F}})/e \approx60\,\mathrm{nm}$ that is shorter than the junction length $L \approx 120\,\mathrm{nm}$. The superconducting coherence length in this diffusive limit is then $\xi_{\mathrm{diff}} \sim \sqrt{\hbar D/\Delta}= 290\,\mathrm{nm}$ \cite{Tafuri2019}, where the diffusion constant $D = v_{\mathrm{F}}l_{\mathrm{e}}/3$. This implies that the junction is in the short diffusive limit ($l_{\mathrm{e}}\ll L \ll \xi_{\mathrm{diff}}$).
\end{appendix}


\begin{thebibliography}{50}%
\makeatletter
\providecommand \@ifxundefined [1]{%
 \@ifx{#1\undefined}
}%
\providecommand \@ifnum [1]{%
 \ifnum #1\expandafter \@firstoftwo
 \else \expandafter \@secondoftwo
 \fi
}%
\providecommand \@ifx [1]{%
 \ifx #1\expandafter \@firstoftwo
 \else \expandafter \@secondoftwo
 \fi
}%
\providecommand \natexlab [1]{#1}%
\providecommand \enquote  [1]{``#1''}%
\providecommand \bibnamefont  [1]{#1}%
\providecommand \bibfnamefont [1]{#1}%
\providecommand \citenamefont [1]{#1}%
\providecommand \href@noop [0]{\@secondoftwo}%
\providecommand \href [0]{\begingroup \@sanitize@url \@href}%
\providecommand \@href[1]{\@@startlink{#1}\@@href}%
\providecommand \@@href[1]{\endgroup#1\@@endlink}%
\providecommand \@sanitize@url [0]{\catcode `\\12\catcode `\$12\catcode
  `\&12\catcode `\#12\catcode `\^12\catcode `\_12\catcode `\%12\relax}%
\providecommand \@@startlink[1]{}%
\providecommand \@@endlink[0]{}%
\providecommand \url  [0]{\begingroup\@sanitize@url \@url }%
\providecommand \@url [1]{\endgroup\@href {#1}{\urlprefix }}%
\providecommand \urlprefix  [0]{URL }%
\providecommand \Eprint [0]{\href }%
\providecommand \doibase [0]{https://doi.org/}%
\providecommand \selectlanguage [0]{\@gobble}%
\providecommand \bibinfo  [0]{\@secondoftwo}%
\providecommand \bibfield  [0]{\@secondoftwo}%
\providecommand \translation [1]{[#1]}%
\providecommand \BibitemOpen [0]{}%
\providecommand \bibitemStop [0]{}%
\providecommand \bibitemNoStop [0]{.\EOS\space}%
\providecommand \EOS [0]{\spacefactor3000\relax}%
\providecommand \BibitemShut  [1]{\csname bibitem#1\endcsname}%
\let\auto@bib@innerbib\@empty
\bibitem [{\citenamefont {Arute}\ \emph {et~al.}(2019)\citenamefont {Arute},
  \citenamefont {Arya}, \citenamefont {Babbush}, \citenamefont {Bacon},
  \citenamefont {Bardin}, \citenamefont {Barends}, \citenamefont {Biswas},
  \citenamefont {Boixo}, \citenamefont {Brandao}, \citenamefont {Buell},
  \citenamefont {Burkett}, \citenamefont {Chen}, \citenamefont {Chen},
  \citenamefont {Chiaro}, \citenamefont {Collins}, \citenamefont {Courtney},
  \citenamefont {Dunsworth}, \citenamefont {Farhi}, \citenamefont {Foxen},
  \citenamefont {Fowler}, \citenamefont {Gidney}, \citenamefont {Giustina},
  \citenamefont {Graff}, \citenamefont {Guerin}, \citenamefont {Habegger},
  \citenamefont {Harrigan}, \citenamefont {Hartmann}, \citenamefont {Ho},
  \citenamefont {Hoffmann}, \citenamefont {Huang}, \citenamefont {Humble},
  \citenamefont {Isakov}, \citenamefont {Jeffrey}, \citenamefont {Jiang},
  \citenamefont {Kafri}, \citenamefont {Kechedzhi}, \citenamefont {Kelly},
  \citenamefont {Klimov}, \citenamefont {Knysh}, \citenamefont {Korotkov},
  \citenamefont {Kostritsa}, \citenamefont {Landhuis}, \citenamefont
  {Lindmark}, \citenamefont {Lucero}, \citenamefont {Lyakh}, \citenamefont
  {Mandrà}, \citenamefont {McClean}, \citenamefont {McEwen}, \citenamefont
  {Megrant}, \citenamefont {Mi}, \citenamefont {Michielsen}, \citenamefont
  {Mohseni}, \citenamefont {Mutus}, \citenamefont {Naaman}, \citenamefont
  {Neeley}, \citenamefont {Neill}, \citenamefont {Niu}, \citenamefont {Ostby},
  \citenamefont {Petukhov}, \citenamefont {Platt}, \citenamefont {Quintana},
  \citenamefont {Rieffel}, \citenamefont {Roushan}, \citenamefont {Rubin},
  \citenamefont {Sank}, \citenamefont {Satzinger}, \citenamefont {Smelyanskiy},
  \citenamefont {Sung}, \citenamefont {Trevithick}, \citenamefont
  {Vainsencher}, \citenamefont {Villalonga}, \citenamefont {White},
  \citenamefont {Yao}, \citenamefont {Yeh}, \citenamefont {Zalcman},
  \citenamefont {Neven},\ and\ \citenamefont {Martinis}}]{Arute2019}%
  \BibitemOpen
  \bibfield  {author} {\bibinfo {author} {\bibfnamefont {F.}~\bibnamefont
  {Arute}}, \bibinfo {author} {\bibfnamefont {K.}~\bibnamefont {Arya}},
  \bibinfo {author} {\bibfnamefont {R.}~\bibnamefont {Babbush}}, \bibinfo
  {author} {\bibfnamefont {D.}~\bibnamefont {Bacon}}, \bibinfo {author}
  {\bibfnamefont {J.~C.}\ \bibnamefont {Bardin}}, \bibinfo {author}
  {\bibfnamefont {R.}~\bibnamefont {Barends}}, \bibinfo {author} {\bibfnamefont
  {R.}~\bibnamefont {Biswas}}, \bibinfo {author} {\bibfnamefont
  {S.}~\bibnamefont {Boixo}}, \bibinfo {author} {\bibfnamefont {F.~G. S.~L.}\
  \bibnamefont {Brandao}}, \bibinfo {author} {\bibfnamefont {D.~A.}\
  \bibnamefont {Buell}}, \bibinfo {author} {\bibfnamefont {B.}~\bibnamefont
  {Burkett}}, \bibinfo {author} {\bibfnamefont {Y.}~\bibnamefont {Chen}},
  \bibinfo {author} {\bibfnamefont {Z.}~\bibnamefont {Chen}}, \bibinfo {author}
  {\bibfnamefont {B.}~\bibnamefont {Chiaro}}, \bibinfo {author} {\bibfnamefont
  {R.}~\bibnamefont {Collins}}, \bibinfo {author} {\bibfnamefont
  {W.}~\bibnamefont {Courtney}}, \bibinfo {author} {\bibfnamefont
  {A.}~\bibnamefont {Dunsworth}}, \bibinfo {author} {\bibfnamefont
  {E.}~\bibnamefont {Farhi}}, \bibinfo {author} {\bibfnamefont
  {B.}~\bibnamefont {Foxen}}, \bibinfo {author} {\bibfnamefont
  {A.}~\bibnamefont {Fowler}}, \bibinfo {author} {\bibfnamefont
  {C.}~\bibnamefont {Gidney}}, \bibinfo {author} {\bibfnamefont
  {M.}~\bibnamefont {Giustina}}, \bibinfo {author} {\bibfnamefont
  {R.}~\bibnamefont {Graff}}, \bibinfo {author} {\bibfnamefont
  {K.}~\bibnamefont {Guerin}}, \bibinfo {author} {\bibfnamefont
  {S.}~\bibnamefont {Habegger}}, \bibinfo {author} {\bibfnamefont {M.~P.}\
  \bibnamefont {Harrigan}}, \bibinfo {author} {\bibfnamefont {M.~J.}\
  \bibnamefont {Hartmann}}, \bibinfo {author} {\bibfnamefont {A.}~\bibnamefont
  {Ho}}, \bibinfo {author} {\bibfnamefont {M.}~\bibnamefont {Hoffmann}},
  \bibinfo {author} {\bibfnamefont {T.}~\bibnamefont {Huang}}, \bibinfo
  {author} {\bibfnamefont {T.~S.}\ \bibnamefont {Humble}}, \bibinfo {author}
  {\bibfnamefont {S.~V.}\ \bibnamefont {Isakov}}, \bibinfo {author}
  {\bibfnamefont {E.}~\bibnamefont {Jeffrey}}, \bibinfo {author} {\bibfnamefont
  {Z.}~\bibnamefont {Jiang}}, \bibinfo {author} {\bibfnamefont
  {D.}~\bibnamefont {Kafri}}, \bibinfo {author} {\bibfnamefont
  {K.}~\bibnamefont {Kechedzhi}}, \bibinfo {author} {\bibfnamefont
  {J.}~\bibnamefont {Kelly}}, \bibinfo {author} {\bibfnamefont {P.~V.}\
  \bibnamefont {Klimov}}, \bibinfo {author} {\bibfnamefont {S.}~\bibnamefont
  {Knysh}}, \bibinfo {author} {\bibfnamefont {A.}~\bibnamefont {Korotkov}},
  \bibinfo {author} {\bibfnamefont {F.}~\bibnamefont {Kostritsa}}, \bibinfo
  {author} {\bibfnamefont {D.}~\bibnamefont {Landhuis}}, \bibinfo {author}
  {\bibfnamefont {M.}~\bibnamefont {Lindmark}}, \bibinfo {author}
  {\bibfnamefont {E.}~\bibnamefont {Lucero}}, \bibinfo {author} {\bibfnamefont
  {D.}~\bibnamefont {Lyakh}}, \bibinfo {author} {\bibfnamefont
  {S.}~\bibnamefont {Mandrà}}, \bibinfo {author} {\bibfnamefont {J.~R.}\
  \bibnamefont {McClean}}, \bibinfo {author} {\bibfnamefont {M.}~\bibnamefont
  {McEwen}}, \bibinfo {author} {\bibfnamefont {A.}~\bibnamefont {Megrant}},
  \bibinfo {author} {\bibfnamefont {X.}~\bibnamefont {Mi}}, \bibinfo {author}
  {\bibfnamefont {K.}~\bibnamefont {Michielsen}}, \bibinfo {author}
  {\bibfnamefont {M.}~\bibnamefont {Mohseni}}, \bibinfo {author} {\bibfnamefont
  {J.}~\bibnamefont {Mutus}}, \bibinfo {author} {\bibfnamefont
  {O.}~\bibnamefont {Naaman}}, \bibinfo {author} {\bibfnamefont
  {M.}~\bibnamefont {Neeley}}, \bibinfo {author} {\bibfnamefont
  {C.}~\bibnamefont {Neill}}, \bibinfo {author} {\bibfnamefont {M.~Y.}\
  \bibnamefont {Niu}}, \bibinfo {author} {\bibfnamefont {E.}~\bibnamefont
  {Ostby}}, \bibinfo {author} {\bibfnamefont {A.}~\bibnamefont {Petukhov}},
  \bibinfo {author} {\bibfnamefont {J.~C.}\ \bibnamefont {Platt}}, \bibinfo
  {author} {\bibfnamefont {C.}~\bibnamefont {Quintana}}, \bibinfo {author}
  {\bibfnamefont {E.~G.}\ \bibnamefont {Rieffel}}, \bibinfo {author}
  {\bibfnamefont {P.}~\bibnamefont {Roushan}}, \bibinfo {author} {\bibfnamefont
  {N.~C.}\ \bibnamefont {Rubin}}, \bibinfo {author} {\bibfnamefont
  {D.}~\bibnamefont {Sank}}, \bibinfo {author} {\bibfnamefont {K.~J.}\
  \bibnamefont {Satzinger}}, \bibinfo {author} {\bibfnamefont {V.}~\bibnamefont
  {Smelyanskiy}}, \bibinfo {author} {\bibfnamefont {K.~J.}\ \bibnamefont
  {Sung}}, \bibinfo {author} {\bibfnamefont {M.~D.}\ \bibnamefont
  {Trevithick}}, \bibinfo {author} {\bibfnamefont {A.}~\bibnamefont
  {Vainsencher}}, \bibinfo {author} {\bibfnamefont {B.}~\bibnamefont
  {Villalonga}}, \bibinfo {author} {\bibfnamefont {T.}~\bibnamefont {White}},
  \bibinfo {author} {\bibfnamefont {Z.~J.}\ \bibnamefont {Yao}}, \bibinfo
  {author} {\bibfnamefont {P.}~\bibnamefont {Yeh}}, \bibinfo {author}
  {\bibfnamefont {A.}~\bibnamefont {Zalcman}}, \bibinfo {author} {\bibfnamefont
  {H.}~\bibnamefont {Neven}},\ and\ \bibinfo {author} {\bibfnamefont {J.~M.}\
  \bibnamefont {Martinis}},\ }\bibfield  {title} {\bibinfo {title} {Quantum
  supremacy using a programmable superconducting processor},\ }\href
  {https://doi.org/10.1038/s41586-019-1666-5} {\bibfield  {journal} {\bibinfo
  {journal} {Nature}\ }\textbf {\bibinfo {volume} {574}},\ \bibinfo {pages}
  {505} (\bibinfo {year} {2019})}\BibitemShut {NoStop}%
\bibitem [{\citenamefont {Reiher}\ \emph {et~al.}(2017)\citenamefont {Reiher},
  \citenamefont {Wiebe}, \citenamefont {Svore}, \citenamefont {Wecker},\ and\
  \citenamefont {Troyer}}]{Reiher2017}%
  \BibitemOpen
  \bibfield  {author} {\bibinfo {author} {\bibfnamefont {M.}~\bibnamefont
  {Reiher}}, \bibinfo {author} {\bibfnamefont {N.}~\bibnamefont {Wiebe}},
  \bibinfo {author} {\bibfnamefont {K.~M.}\ \bibnamefont {Svore}}, \bibinfo
  {author} {\bibfnamefont {D.}~\bibnamefont {Wecker}},\ and\ \bibinfo {author}
  {\bibfnamefont {M.}~\bibnamefont {Troyer}},\ }\bibfield  {title} {\bibinfo
  {title} {Elucidating reaction mechanisms on quantum computers},\ }\href
  {https://doi.org/10.1073/pnas.1619152114} {\bibfield  {journal} {\bibinfo
  {journal} {Proc. Natl. Acad. Sci. U.S.A}\ }\textbf {\bibinfo {volume}
  {114}},\ \bibinfo {pages} {7555} (\bibinfo {year} {2017})}\BibitemShut
  {NoStop}%
\bibitem [{\citenamefont {Tahan}(2019)}]{Tahan2019}%
  \BibitemOpen
  \bibfield  {author} {\bibinfo {author} {\bibfnamefont {C.}~\bibnamefont
  {Tahan}},\ }\bibfield  {title} {\bibinfo {title} {Graphene qubit motivates
  materials science},\ }\href {https://doi.org/10.1038/s41565-019-0369-2}
  {\bibfield  {journal} {\bibinfo  {journal} {Nat. Nanotechnol.}\ }\textbf
  {\bibinfo {volume} {14}},\ \bibinfo {pages} {102} (\bibinfo {year}
  {2019})}\BibitemShut {NoStop}%
\bibitem [{\citenamefont {Larsen}\ \emph {et~al.}(2015)\citenamefont {Larsen},
  \citenamefont {Petersson}, \citenamefont {Kuemmeth}, \citenamefont
  {Jespersen}, \citenamefont {Krogstrup}, \citenamefont {Nyg\aa{}rd},\ and\
  \citenamefont {Marcus}}]{Larsen2015}%
  \BibitemOpen
  \bibfield  {author} {\bibinfo {author} {\bibfnamefont {T.~W.}\ \bibnamefont
  {Larsen}}, \bibinfo {author} {\bibfnamefont {K.~D.}\ \bibnamefont
  {Petersson}}, \bibinfo {author} {\bibfnamefont {F.}~\bibnamefont {Kuemmeth}},
  \bibinfo {author} {\bibfnamefont {T.~S.}\ \bibnamefont {Jespersen}}, \bibinfo
  {author} {\bibfnamefont {P.}~\bibnamefont {Krogstrup}}, \bibinfo {author}
  {\bibfnamefont {J.}~\bibnamefont {Nyg\aa{}rd}},\ and\ \bibinfo {author}
  {\bibfnamefont {C.~M.}\ \bibnamefont {Marcus}},\ }\bibfield  {title}
  {\bibinfo {title} {{Semiconductor-Nanowire-Based Superconducting Qubit}},\
  }\href {https://doi.org/10.1103/PhysRevLett.115.127001} {\bibfield  {journal}
  {\bibinfo  {journal} {Phys. Rev. Lett.}\ }\textbf {\bibinfo {volume} {115}},\
  \bibinfo {pages} {127001} (\bibinfo {year} {2015})}\BibitemShut {NoStop}%
\bibitem [{\citenamefont {Casparis}\ \emph {et~al.}(2018)\citenamefont
  {Casparis}, \citenamefont {Connolly}, \citenamefont {Kjaergaard},
  \citenamefont {Pearson}, \citenamefont {Kringh{\o}j}, \citenamefont {Larsen},
  \citenamefont {Kuemmeth}, \citenamefont {Wang}, \citenamefont {Thomas},
  \citenamefont {Gronin}, \citenamefont {Gardner}, \citenamefont {Manfra},
  \citenamefont {Marcus},\ and\ \citenamefont {Petersson}}]{Casparis2018}%
  \BibitemOpen
  \bibfield  {author} {\bibinfo {author} {\bibfnamefont {L.}~\bibnamefont
  {Casparis}}, \bibinfo {author} {\bibfnamefont {M.~R.}\ \bibnamefont
  {Connolly}}, \bibinfo {author} {\bibfnamefont {M.}~\bibnamefont
  {Kjaergaard}}, \bibinfo {author} {\bibfnamefont {N.~J.}\ \bibnamefont
  {Pearson}}, \bibinfo {author} {\bibfnamefont {A.}~\bibnamefont
  {Kringh{\o}j}}, \bibinfo {author} {\bibfnamefont {T.~W.}\ \bibnamefont
  {Larsen}}, \bibinfo {author} {\bibfnamefont {F.}~\bibnamefont {Kuemmeth}},
  \bibinfo {author} {\bibfnamefont {T.}~\bibnamefont {Wang}}, \bibinfo {author}
  {\bibfnamefont {C.}~\bibnamefont {Thomas}}, \bibinfo {author} {\bibfnamefont
  {S.}~\bibnamefont {Gronin}}, \bibinfo {author} {\bibfnamefont {G.~C.}\
  \bibnamefont {Gardner}}, \bibinfo {author} {\bibfnamefont {M.~J.}\
  \bibnamefont {Manfra}}, \bibinfo {author} {\bibfnamefont {C.~M.}\
  \bibnamefont {Marcus}},\ and\ \bibinfo {author} {\bibfnamefont {K.~D.}\
  \bibnamefont {Petersson}},\ }\bibfield  {title} {\bibinfo {title}
  {{Superconducting gatemon qubit based on a proximitized two-dimensional
  electron gas}},\ }\href {https://doi.org/10.1038/s41565-018-0207-y}
  {\bibfield  {journal} {\bibinfo  {journal} {Nat. Nanotechnol.}\ }\textbf
  {\bibinfo {volume} {13}},\ \bibinfo {pages} {915} (\bibinfo {year}
  {2018})}\BibitemShut {NoStop}%
\bibitem [{\citenamefont {Wang}\ \emph {et~al.}(2019)\citenamefont {Wang},
  \citenamefont {Rodan-Legrain}, \citenamefont {Bretheau}, \citenamefont
  {Campbell}, \citenamefont {Kannan}, \citenamefont {Kim}, \citenamefont
  {Kjaergaard}, \citenamefont {Krantz}, \citenamefont {Samach}, \citenamefont
  {Yan}, \citenamefont {Yoder}, \citenamefont {Watanabe}, \citenamefont
  {Taniguchi}, \citenamefont {Orlando}, \citenamefont {Gustavsson},
  \citenamefont {Jarillo-Herrero},\ and\ \citenamefont {Oliver}}]{Wang2019}%
  \BibitemOpen
  \bibfield  {author} {\bibinfo {author} {\bibfnamefont {J.~I.-J.}\
  \bibnamefont {Wang}}, \bibinfo {author} {\bibfnamefont {D.}~\bibnamefont
  {Rodan-Legrain}}, \bibinfo {author} {\bibfnamefont {L.}~\bibnamefont
  {Bretheau}}, \bibinfo {author} {\bibfnamefont {D.~L.}\ \bibnamefont
  {Campbell}}, \bibinfo {author} {\bibfnamefont {B.}~\bibnamefont {Kannan}},
  \bibinfo {author} {\bibfnamefont {D.}~\bibnamefont {Kim}}, \bibinfo {author}
  {\bibfnamefont {M.}~\bibnamefont {Kjaergaard}}, \bibinfo {author}
  {\bibfnamefont {P.}~\bibnamefont {Krantz}}, \bibinfo {author} {\bibfnamefont
  {G.~O.}\ \bibnamefont {Samach}}, \bibinfo {author} {\bibfnamefont
  {F.}~\bibnamefont {Yan}}, \bibinfo {author} {\bibfnamefont {J.~L.}\
  \bibnamefont {Yoder}}, \bibinfo {author} {\bibfnamefont {K.}~\bibnamefont
  {Watanabe}}, \bibinfo {author} {\bibfnamefont {T.}~\bibnamefont {Taniguchi}},
  \bibinfo {author} {\bibfnamefont {T.~P.}\ \bibnamefont {Orlando}}, \bibinfo
  {author} {\bibfnamefont {S.}~\bibnamefont {Gustavsson}}, \bibinfo {author}
  {\bibfnamefont {P.}~\bibnamefont {Jarillo-Herrero}},\ and\ \bibinfo {author}
  {\bibfnamefont {W.~D.}\ \bibnamefont {Oliver}},\ }\bibfield  {title}
  {\bibinfo {title} {{Coherent control of a hybrid superconducting circuit made
  with graphene-based van der Waals heterostructures}},\ }\href
  {https://doi.org/10.1038/s41565-018-0329-2} {\bibfield  {journal} {\bibinfo
  {journal} {Nat. Nanotechnol.}\ }\textbf {\bibinfo {volume} {14}},\ \bibinfo
  {pages} {120} (\bibinfo {year} {2019})}\BibitemShut {NoStop}%
\bibitem [{\citenamefont {Pauka}\ \emph {et~al.}(2019)\citenamefont {Pauka},
  \citenamefont {Das}, \citenamefont {Kalra}, \citenamefont {Moini},
  \citenamefont {Yang}, \citenamefont {Trainer}, \citenamefont {Bousquet},
  \citenamefont {Cantaloube}, \citenamefont {Dick}, \citenamefont {Gardner},
  \citenamefont {Manfra},\ and\ \citenamefont {Reilly}}]{Pauka2019}%
  \BibitemOpen
  \bibfield  {author} {\bibinfo {author} {\bibfnamefont {S.~J.}\ \bibnamefont
  {Pauka}}, \bibinfo {author} {\bibfnamefont {K.}~\bibnamefont {Das}}, \bibinfo
  {author} {\bibfnamefont {R.}~\bibnamefont {Kalra}}, \bibinfo {author}
  {\bibfnamefont {A.}~\bibnamefont {Moini}}, \bibinfo {author} {\bibfnamefont
  {Y.}~\bibnamefont {Yang}}, \bibinfo {author} {\bibfnamefont {M.}~\bibnamefont
  {Trainer}}, \bibinfo {author} {\bibfnamefont {A.}~\bibnamefont {Bousquet}},
  \bibinfo {author} {\bibfnamefont {C.}~\bibnamefont {Cantaloube}}, \bibinfo
  {author} {\bibfnamefont {N.}~\bibnamefont {Dick}}, \bibinfo {author}
  {\bibfnamefont {G.~C.}\ \bibnamefont {Gardner}}, \bibinfo {author}
  {\bibfnamefont {M.~J.}\ \bibnamefont {Manfra}},\ and\ \bibinfo {author}
  {\bibfnamefont {D.~J.}\ \bibnamefont {Reilly}},\ }\bibfield  {title}
  {\bibinfo {title} {{A Cryogenic Interface for Controlling Many Qubits}},\
  }\href {https://arxiv.org/abs/1912.01299} {\bibfield  {journal} {\bibinfo
  {journal} {arXiv:1912.01299}\ } (\bibinfo {year} {2019})}\BibitemShut
  {NoStop}%
\bibitem [{\citenamefont {Krizek}\ \emph {et~al.}(2018)\citenamefont {Krizek},
  \citenamefont {Sestoft}, \citenamefont {Aseev}, \citenamefont
  {Marti-Sanchez}, \citenamefont {Vaitiek\ifmmode~\dot{e}\else \.{e}\fi{}nas},
  \citenamefont {Casparis}, \citenamefont {Khan}, \citenamefont {Liu},
  \citenamefont {Stankevi\ifmmode~\check{c}\else \v{c}\fi{}}, \citenamefont
  {Whiticar}, \citenamefont {Fursina}, \citenamefont {Boekhout}, \citenamefont
  {Koops}, \citenamefont {Uccelli}, \citenamefont {Kouwenhoven}, \citenamefont
  {Marcus}, \citenamefont {Arbiol},\ and\ \citenamefont
  {Krogstrup}}]{Krizek2018}%
  \BibitemOpen
  \bibfield  {author} {\bibinfo {author} {\bibfnamefont {F.}~\bibnamefont
  {Krizek}}, \bibinfo {author} {\bibfnamefont {J.~E.}\ \bibnamefont {Sestoft}},
  \bibinfo {author} {\bibfnamefont {P.}~\bibnamefont {Aseev}}, \bibinfo
  {author} {\bibfnamefont {S.}~\bibnamefont {Marti-Sanchez}}, \bibinfo {author}
  {\bibfnamefont {S.}~\bibnamefont {Vaitiek\ifmmode~\dot{e}\else
  \.{e}\fi{}nas}}, \bibinfo {author} {\bibfnamefont {L.}~\bibnamefont
  {Casparis}}, \bibinfo {author} {\bibfnamefont {S.~A.}\ \bibnamefont {Khan}},
  \bibinfo {author} {\bibfnamefont {Y.}~\bibnamefont {Liu}}, \bibinfo {author}
  {\bibfnamefont {T.~c.~v.}\ \bibnamefont {Stankevi\ifmmode~\check{c}\else
  \v{c}\fi{}}}, \bibinfo {author} {\bibfnamefont {A.~M.}\ \bibnamefont
  {Whiticar}}, \bibinfo {author} {\bibfnamefont {A.}~\bibnamefont {Fursina}},
  \bibinfo {author} {\bibfnamefont {F.}~\bibnamefont {Boekhout}}, \bibinfo
  {author} {\bibfnamefont {R.}~\bibnamefont {Koops}}, \bibinfo {author}
  {\bibfnamefont {E.}~\bibnamefont {Uccelli}}, \bibinfo {author} {\bibfnamefont
  {L.~P.}\ \bibnamefont {Kouwenhoven}}, \bibinfo {author} {\bibfnamefont
  {C.~M.}\ \bibnamefont {Marcus}}, \bibinfo {author} {\bibfnamefont
  {J.}~\bibnamefont {Arbiol}},\ and\ \bibinfo {author} {\bibfnamefont
  {P.}~\bibnamefont {Krogstrup}},\ }\bibfield  {title} {\bibinfo {title}
  {{Field effect enhancement in buffered quantum nanowire networks}},\ }\href
  {https://doi.org/10.1103/PhysRevMaterials.2.093401} {\bibfield  {journal}
  {\bibinfo  {journal} {Phys. Rev. Mater.}\ }\textbf {\bibinfo {volume} {2}},\
  \bibinfo {pages} {093401} (\bibinfo {year} {2018})}\BibitemShut {NoStop}%
\bibitem [{\citenamefont {Chang}\ \emph {et~al.}(2015)\citenamefont {Chang},
  \citenamefont {Albrecht}, \citenamefont {Jespersen}, \citenamefont
  {Kuemmeth}, \citenamefont {Krogstrup}, \citenamefont {Nyg{\aa}rd},\ and\
  \citenamefont {Marcus}}]{Chang2015}%
  \BibitemOpen
  \bibfield  {author} {\bibinfo {author} {\bibfnamefont {W.}~\bibnamefont
  {Chang}}, \bibinfo {author} {\bibfnamefont {S.~M.}\ \bibnamefont {Albrecht}},
  \bibinfo {author} {\bibfnamefont {T.~S.}\ \bibnamefont {Jespersen}}, \bibinfo
  {author} {\bibfnamefont {F.}~\bibnamefont {Kuemmeth}}, \bibinfo {author}
  {\bibfnamefont {P.}~\bibnamefont {Krogstrup}}, \bibinfo {author}
  {\bibfnamefont {J.}~\bibnamefont {Nyg{\aa}rd}},\ and\ \bibinfo {author}
  {\bibfnamefont {C.~M.}\ \bibnamefont {Marcus}},\ }\bibfield  {title}
  {\bibinfo {title} {{Hard gap in epitaxial semiconductor--superconductor
  nanowires}},\ }\href {https://doi.org/10.1038/nnano.2014.306} {\bibfield
  {journal} {\bibinfo  {journal} {Nat. Nanotechnol.}\ }\textbf {\bibinfo
  {volume} {10}},\ \bibinfo {pages} {232} (\bibinfo {year} {2015})}\BibitemShut
  {NoStop}%
\bibitem [{\citenamefont {Bilmes}\ \emph {et~al.}(2017)\citenamefont {Bilmes},
  \citenamefont {Zanker}, \citenamefont {Heimes}, \citenamefont {Marthaler},
  \citenamefont {Sch\"on}, \citenamefont {Weiss}, \citenamefont {Ustinov},\
  and\ \citenamefont {Lisenfeld}}]{Bilmes2017}%
  \BibitemOpen
  \bibfield  {author} {\bibinfo {author} {\bibfnamefont {A.}~\bibnamefont
  {Bilmes}}, \bibinfo {author} {\bibfnamefont {S.}~\bibnamefont {Zanker}},
  \bibinfo {author} {\bibfnamefont {A.}~\bibnamefont {Heimes}}, \bibinfo
  {author} {\bibfnamefont {M.}~\bibnamefont {Marthaler}}, \bibinfo {author}
  {\bibfnamefont {G.}~\bibnamefont {Sch\"on}}, \bibinfo {author} {\bibfnamefont
  {G.}~\bibnamefont {Weiss}}, \bibinfo {author} {\bibfnamefont {A.~V.}\
  \bibnamefont {Ustinov}},\ and\ \bibinfo {author} {\bibfnamefont
  {J.}~\bibnamefont {Lisenfeld}},\ }\bibfield  {title} {\bibinfo {title}
  {{Electronic decoherence of two-level systems in a Josephson junction}},\
  }\href {https://doi.org/10.1103/PhysRevB.96.064504} {\bibfield  {journal}
  {\bibinfo  {journal} {Phys. Rev. B}\ }\textbf {\bibinfo {volume} {96}},\
  \bibinfo {pages} {064504} (\bibinfo {year} {2017})}\BibitemShut {NoStop}%
\bibitem [{\citenamefont {Casparis}\ \emph {et~al.}(2016)\citenamefont
  {Casparis}, \citenamefont {Larsen}, \citenamefont {Olsen}, \citenamefont
  {Kuemmeth}, \citenamefont {Krogstrup}, \citenamefont {Nyg\aa{}rd},
  \citenamefont {Petersson},\ and\ \citenamefont {Marcus}}]{Casparis2016}%
  \BibitemOpen
  \bibfield  {author} {\bibinfo {author} {\bibfnamefont {L.}~\bibnamefont
  {Casparis}}, \bibinfo {author} {\bibfnamefont {T.~W.}\ \bibnamefont
  {Larsen}}, \bibinfo {author} {\bibfnamefont {M.~S.}\ \bibnamefont {Olsen}},
  \bibinfo {author} {\bibfnamefont {F.}~\bibnamefont {Kuemmeth}}, \bibinfo
  {author} {\bibfnamefont {P.}~\bibnamefont {Krogstrup}}, \bibinfo {author}
  {\bibfnamefont {J.}~\bibnamefont {Nyg\aa{}rd}}, \bibinfo {author}
  {\bibfnamefont {K.~D.}\ \bibnamefont {Petersson}},\ and\ \bibinfo {author}
  {\bibfnamefont {C.~M.}\ \bibnamefont {Marcus}},\ }\bibfield  {title}
  {\bibinfo {title} {{Gatemon Benchmarking and Two-Qubit Operations}},\ }\href
  {https://doi.org/10.1103/PhysRevLett.116.150505} {\bibfield  {journal}
  {\bibinfo  {journal} {Phys. Rev. Lett.}\ }\textbf {\bibinfo {volume} {116}},\
  \bibinfo {pages} {150505} (\bibinfo {year} {2016})}\BibitemShut {NoStop}%
\bibitem [{\citenamefont {Luthi}\ \emph {et~al.}(2018)\citenamefont {Luthi},
  \citenamefont {Stavenga}, \citenamefont {Enzing}, \citenamefont {Bruno},
  \citenamefont {Dickel}, \citenamefont {Langford}, \citenamefont {Rol},
  \citenamefont {Jespersen}, \citenamefont {Nyg\aa{}rd}, \citenamefont
  {Krogstrup},\ and\ \citenamefont {DiCarlo}}]{Luthi2018}%
  \BibitemOpen
  \bibfield  {author} {\bibinfo {author} {\bibfnamefont {F.}~\bibnamefont
  {Luthi}}, \bibinfo {author} {\bibfnamefont {T.}~\bibnamefont {Stavenga}},
  \bibinfo {author} {\bibfnamefont {O.~W.}\ \bibnamefont {Enzing}}, \bibinfo
  {author} {\bibfnamefont {A.}~\bibnamefont {Bruno}}, \bibinfo {author}
  {\bibfnamefont {C.}~\bibnamefont {Dickel}}, \bibinfo {author} {\bibfnamefont
  {N.~K.}\ \bibnamefont {Langford}}, \bibinfo {author} {\bibfnamefont {M.~A.}\
  \bibnamefont {Rol}}, \bibinfo {author} {\bibfnamefont {T.~S.}\ \bibnamefont
  {Jespersen}}, \bibinfo {author} {\bibfnamefont {J.}~\bibnamefont
  {Nyg\aa{}rd}}, \bibinfo {author} {\bibfnamefont {P.}~\bibnamefont
  {Krogstrup}},\ and\ \bibinfo {author} {\bibfnamefont {L.}~\bibnamefont
  {DiCarlo}},\ }\bibfield  {title} {\bibinfo {title} {{Evolution of Nanowire
  Transmon Qubits and Their Coherence in a Magnetic Field}},\ }\href
  {https://doi.org/10.1103/PhysRevLett.120.100502} {\bibfield  {journal}
  {\bibinfo  {journal} {Phys. Rev. Lett.}\ }\textbf {\bibinfo {volume} {120}},\
  \bibinfo {pages} {100502} (\bibinfo {year} {2018})}\BibitemShut {NoStop}%
\bibitem [{\citenamefont {Dunsworth}\ \emph {et~al.}(2017)\citenamefont
  {Dunsworth}, \citenamefont {Megrant}, \citenamefont {Quintana}, \citenamefont
  {Chen}, \citenamefont {Barends}, \citenamefont {Burkett}, \citenamefont
  {Foxen}, \citenamefont {Chen}, \citenamefont {Chiaro}, \citenamefont
  {Fowler}, \citenamefont {Graff}, \citenamefont {Jeffrey}, \citenamefont
  {Kelly}, \citenamefont {Lucero}, \citenamefont {Mutus}, \citenamefont
  {Neeley}, \citenamefont {Neill}, \citenamefont {Roushan}, \citenamefont
  {Sank}, \citenamefont {Vainsencher}, \citenamefont {Wenner}, \citenamefont
  {White},\ and\ \citenamefont {Martinis}}]{Dunsworth2017}%
  \BibitemOpen
  \bibfield  {author} {\bibinfo {author} {\bibfnamefont {A.}~\bibnamefont
  {Dunsworth}}, \bibinfo {author} {\bibfnamefont {A.}~\bibnamefont {Megrant}},
  \bibinfo {author} {\bibfnamefont {C.}~\bibnamefont {Quintana}}, \bibinfo
  {author} {\bibfnamefont {Z.}~\bibnamefont {Chen}}, \bibinfo {author}
  {\bibfnamefont {R.}~\bibnamefont {Barends}}, \bibinfo {author} {\bibfnamefont
  {B.}~\bibnamefont {Burkett}}, \bibinfo {author} {\bibfnamefont
  {B.}~\bibnamefont {Foxen}}, \bibinfo {author} {\bibfnamefont
  {Y.}~\bibnamefont {Chen}}, \bibinfo {author} {\bibfnamefont {B.}~\bibnamefont
  {Chiaro}}, \bibinfo {author} {\bibfnamefont {A.}~\bibnamefont {Fowler}},
  \bibinfo {author} {\bibfnamefont {R.}~\bibnamefont {Graff}}, \bibinfo
  {author} {\bibfnamefont {E.}~\bibnamefont {Jeffrey}}, \bibinfo {author}
  {\bibfnamefont {J.}~\bibnamefont {Kelly}}, \bibinfo {author} {\bibfnamefont
  {E.}~\bibnamefont {Lucero}}, \bibinfo {author} {\bibfnamefont {J.~Y.}\
  \bibnamefont {Mutus}}, \bibinfo {author} {\bibfnamefont {M.}~\bibnamefont
  {Neeley}}, \bibinfo {author} {\bibfnamefont {C.}~\bibnamefont {Neill}},
  \bibinfo {author} {\bibfnamefont {P.}~\bibnamefont {Roushan}}, \bibinfo
  {author} {\bibfnamefont {D.}~\bibnamefont {Sank}}, \bibinfo {author}
  {\bibfnamefont {A.}~\bibnamefont {Vainsencher}}, \bibinfo {author}
  {\bibfnamefont {J.}~\bibnamefont {Wenner}}, \bibinfo {author} {\bibfnamefont
  {T.~C.}\ \bibnamefont {White}},\ and\ \bibinfo {author} {\bibfnamefont
  {J.~M.}\ \bibnamefont {Martinis}},\ }\bibfield  {title} {\bibinfo {title}
  {{Characterization and reduction of capacitive loss induced by sub-micron
  Josephson junction fabrication in superconducting qubits}},\ }\href
  {https://doi.org/10.1063/1.4993577} {\bibfield  {journal} {\bibinfo
  {journal} {Appl. Phys. Lett.}\ }\textbf {\bibinfo {volume} {111}},\ \bibinfo
  {pages} {022601} (\bibinfo {year} {2017})}\BibitemShut {NoStop}%
\bibitem [{\citenamefont {Burnett}\ \emph {et~al.}(2019)\citenamefont
  {Burnett}, \citenamefont {Bengtsson}, \citenamefont {Scigliuzzo},
  \citenamefont {Niepce}, \citenamefont {Kudra}, \citenamefont {Delsing},\ and\
  \citenamefont {Bylander}}]{Burnett2019}%
  \BibitemOpen
  \bibfield  {author} {\bibinfo {author} {\bibfnamefont {J.~J.}\ \bibnamefont
  {Burnett}}, \bibinfo {author} {\bibfnamefont {A.}~\bibnamefont {Bengtsson}},
  \bibinfo {author} {\bibfnamefont {M.}~\bibnamefont {Scigliuzzo}}, \bibinfo
  {author} {\bibfnamefont {D.}~\bibnamefont {Niepce}}, \bibinfo {author}
  {\bibfnamefont {M.}~\bibnamefont {Kudra}}, \bibinfo {author} {\bibfnamefont
  {P.}~\bibnamefont {Delsing}},\ and\ \bibinfo {author} {\bibfnamefont
  {J.}~\bibnamefont {Bylander}},\ }\bibfield  {title} {\bibinfo {title}
  {Decoherence benchmarking of superconducting qubits},\ }\href
  {https://doi.org/10.1038/s41534-019-0168-5} {\bibfield  {journal} {\bibinfo
  {journal} {npj Quantum Inf.}\ }\textbf {\bibinfo {volume} {5}},\ \bibinfo
  {pages} {54} (\bibinfo {year} {2019})}\BibitemShut {NoStop}%
\bibitem [{\citenamefont {N{\'{e}}meth}\ \emph {et~al.}(2008)\citenamefont
  {N{\'{e}}meth}, \citenamefont {Kunert}, \citenamefont {Stolz},\ and\
  \citenamefont {Volz}}]{Nemeth2008}%
  \BibitemOpen
  \bibfield  {author} {\bibinfo {author} {\bibfnamefont {I.}~\bibnamefont
  {N{\'{e}}meth}}, \bibinfo {author} {\bibfnamefont {B.}~\bibnamefont
  {Kunert}}, \bibinfo {author} {\bibfnamefont {W.}~\bibnamefont {Stolz}},\ and\
  \bibinfo {author} {\bibfnamefont {K.}~\bibnamefont {Volz}},\ }\bibfield
  {title} {\bibinfo {title} {{Heteroepitaxy of {GaP} on Si: Correlation of
  morphology, anti-phase-domain structure and {MOVPE} growth conditions}},\
  }\href {https://doi.org/10.1016/j.jcrysgro.2007.11.127} {\bibfield  {journal}
  {\bibinfo  {journal} {J. Cryst. Growth}\ }\textbf {\bibinfo {volume} {310}},\
  \bibinfo {pages} {1595} (\bibinfo {year} {2008})}\BibitemShut {NoStop}%
\bibitem [{\citenamefont {Döscher}\ \emph {et~al.}(2008)\citenamefont
  {Döscher}, \citenamefont {Hannappel}, \citenamefont {Kunert}, \citenamefont
  {Beyer}, \citenamefont {Volz},\ and\ \citenamefont {Stolz}}]{Doescher2008}%
  \BibitemOpen
  \bibfield  {author} {\bibinfo {author} {\bibfnamefont {H.}~\bibnamefont
  {Döscher}}, \bibinfo {author} {\bibfnamefont {T.}~\bibnamefont {Hannappel}},
  \bibinfo {author} {\bibfnamefont {B.}~\bibnamefont {Kunert}}, \bibinfo
  {author} {\bibfnamefont {A.}~\bibnamefont {Beyer}}, \bibinfo {author}
  {\bibfnamefont {K.}~\bibnamefont {Volz}},\ and\ \bibinfo {author}
  {\bibfnamefont {W.}~\bibnamefont {Stolz}},\ }\bibfield  {title} {\bibinfo
  {title} {{In situ verification of single-domain {III}-{V} on {Si}(100) growth
  via metal-organic vapor phase epitaxy}},\ }\href
  {https://doi.org/10.1063/1.3009570} {\bibfield  {journal} {\bibinfo
  {journal} {Appl. Phys. Lett.}\ }\textbf {\bibinfo {volume} {93}},\ \bibinfo
  {pages} {172110} (\bibinfo {year} {2008})}\BibitemShut {NoStop}%
\bibitem [{\citenamefont {Gardner}\ \emph {et~al.}(2016)\citenamefont
  {Gardner}, \citenamefont {Fallahi}, \citenamefont {Watson},\ and\
  \citenamefont {Manfra}}]{Gardner2016}%
  \BibitemOpen
  \bibfield  {author} {\bibinfo {author} {\bibfnamefont {G.~C.}\ \bibnamefont
  {Gardner}}, \bibinfo {author} {\bibfnamefont {S.}~\bibnamefont {Fallahi}},
  \bibinfo {author} {\bibfnamefont {J.~D.}\ \bibnamefont {Watson}},\ and\
  \bibinfo {author} {\bibfnamefont {M.~J.}\ \bibnamefont {Manfra}},\ }\bibfield
   {title} {\bibinfo {title} {{Modified MBE hardware and techniques and role of
  gallium purity for attainment of two dimensional electron gas mobility
  $>35\times10^6\mathrm{cm^2/Vs}$ in AlGaAs/GaAs quantum wells grown by MBE}},\
  }\href {https://doi.org/https://doi.org/10.1016/j.jcrysgro.2016.02.010}
  {\bibfield  {journal} {\bibinfo  {journal} {J. Cryst. Growth}\ }\textbf
  {\bibinfo {volume} {441}},\ \bibinfo {pages} {71} (\bibinfo {year}
  {2016})}\BibitemShut {NoStop}%
\bibitem [{\citenamefont {Krogstrup}\ \emph {et~al.}(2015)\citenamefont
  {Krogstrup}, \citenamefont {Ziino}, \citenamefont {Chang}, \citenamefont
  {Albrecht}, \citenamefont {Madsen}, \citenamefont {Johnson}, \citenamefont
  {Nyg{\aa}rd}, \citenamefont {Marcus},\ and\ \citenamefont
  {Jespersen}}]{Krogstrup2015}%
  \BibitemOpen
  \bibfield  {author} {\bibinfo {author} {\bibfnamefont {P.}~\bibnamefont
  {Krogstrup}}, \bibinfo {author} {\bibfnamefont {N.~L.~B.}\ \bibnamefont
  {Ziino}}, \bibinfo {author} {\bibfnamefont {W.}~\bibnamefont {Chang}},
  \bibinfo {author} {\bibfnamefont {S.~M.}\ \bibnamefont {Albrecht}}, \bibinfo
  {author} {\bibfnamefont {M.~H.}\ \bibnamefont {Madsen}}, \bibinfo {author}
  {\bibfnamefont {E.}~\bibnamefont {Johnson}}, \bibinfo {author} {\bibfnamefont
  {J.}~\bibnamefont {Nyg{\aa}rd}}, \bibinfo {author} {\bibfnamefont {C.~M.}\
  \bibnamefont {Marcus}},\ and\ \bibinfo {author} {\bibfnamefont {T.~S.}\
  \bibnamefont {Jespersen}},\ }\bibfield  {title} {\bibinfo {title} {Epitaxy of
  semiconductor--superconductor nanowires},\ }\href
  {https://doi.org/10.1038/nmat4176} {\bibfield  {journal} {\bibinfo  {journal}
  {Nat. Mater.}\ }\textbf {\bibinfo {volume} {14}},\ \bibinfo {pages} {400}
  (\bibinfo {year} {2015})}\BibitemShut {NoStop}%
\bibitem [{\citenamefont {Aseev}\ \emph {et~al.}(2019)\citenamefont {Aseev},
  \citenamefont {Fursina}, \citenamefont {Boekhout}, \citenamefont {Krizek},
  \citenamefont {Sestoft}, \citenamefont {Borsoi}, \citenamefont {Heedt},
  \citenamefont {Wang}, \citenamefont {Binci}, \citenamefont
  {Mart{\'\i}-S{\'a}nchez}, \citenamefont {Swoboda}, \citenamefont {Koops},
  \citenamefont {Uccelli}, \citenamefont {Arbiol}, \citenamefont {Krogstrup},
  \citenamefont {Kouwenhoven},\ and\ \citenamefont {Caroff}}]{Aseev2019}%
  \BibitemOpen
  \bibfield  {author} {\bibinfo {author} {\bibfnamefont {P.}~\bibnamefont
  {Aseev}}, \bibinfo {author} {\bibfnamefont {A.}~\bibnamefont {Fursina}},
  \bibinfo {author} {\bibfnamefont {F.}~\bibnamefont {Boekhout}}, \bibinfo
  {author} {\bibfnamefont {F.}~\bibnamefont {Krizek}}, \bibinfo {author}
  {\bibfnamefont {J.~E.}\ \bibnamefont {Sestoft}}, \bibinfo {author}
  {\bibfnamefont {F.}~\bibnamefont {Borsoi}}, \bibinfo {author} {\bibfnamefont
  {S.}~\bibnamefont {Heedt}}, \bibinfo {author} {\bibfnamefont
  {G.}~\bibnamefont {Wang}}, \bibinfo {author} {\bibfnamefont {L.}~\bibnamefont
  {Binci}}, \bibinfo {author} {\bibfnamefont {S.}~\bibnamefont
  {Mart{\'\i}-S{\'a}nchez}}, \bibinfo {author} {\bibfnamefont {T.}~\bibnamefont
  {Swoboda}}, \bibinfo {author} {\bibfnamefont {R.}~\bibnamefont {Koops}},
  \bibinfo {author} {\bibfnamefont {E.}~\bibnamefont {Uccelli}}, \bibinfo
  {author} {\bibfnamefont {J.}~\bibnamefont {Arbiol}}, \bibinfo {author}
  {\bibfnamefont {P.}~\bibnamefont {Krogstrup}}, \bibinfo {author}
  {\bibfnamefont {L.~P.}\ \bibnamefont {Kouwenhoven}},\ and\ \bibinfo {author}
  {\bibfnamefont {P.}~\bibnamefont {Caroff}},\ }\bibfield  {title} {\bibinfo
  {title} {{Selectivity Map for Molecular Beam Epitaxy of Advanced III--V
  Quantum Nanowire Networks}},\ }\bibfield  {booktitle} {\emph {\bibinfo
  {booktitle} {Nano Lett.}},\ }\href
  {https://doi.org/10.1021/acs.nanolett.8b03733} {\bibfield  {journal}
  {\bibinfo  {journal} {Nano Lett.}\ }\textbf {\bibinfo {volume} {19}},\
  \bibinfo {pages} {218} (\bibinfo {year} {2019})}\BibitemShut {NoStop}%
\bibitem [{\citenamefont {G{\"u}nel}\ \emph {et~al.}(2014)\citenamefont
  {G{\"u}nel}, \citenamefont {Borgwardt}, \citenamefont {Batov}, \citenamefont
  {Hardtdegen}, \citenamefont {Sladek}, \citenamefont {Panaitov}, \citenamefont
  {Gr{\"u}tzmacher},\ and\ \citenamefont {Sch{\"a}pers}}]{Guenel2014}%
  \BibitemOpen
  \bibfield  {author} {\bibinfo {author} {\bibfnamefont {H.~Y.}\ \bibnamefont
  {G{\"u}nel}}, \bibinfo {author} {\bibfnamefont {N.}~\bibnamefont
  {Borgwardt}}, \bibinfo {author} {\bibfnamefont {I.~E.}\ \bibnamefont
  {Batov}}, \bibinfo {author} {\bibfnamefont {H.}~\bibnamefont {Hardtdegen}},
  \bibinfo {author} {\bibfnamefont {K.}~\bibnamefont {Sladek}}, \bibinfo
  {author} {\bibfnamefont {G.}~\bibnamefont {Panaitov}}, \bibinfo {author}
  {\bibfnamefont {D.}~\bibnamefont {Gr{\"u}tzmacher}},\ and\ \bibinfo {author}
  {\bibfnamefont {T.}~\bibnamefont {Sch{\"a}pers}},\ }\bibfield  {title}
  {\bibinfo {title} {{Crossover from Josephson Effect to Single Interface
  Andreev Reflection in Asymmetric Superconductor/Nanowire Junctions}},\ }\href
  {https://doi.org/10.1021/nl501350v} {\bibfield  {journal} {\bibinfo
  {journal} {Nano Lett.}\ }\textbf {\bibinfo {volume} {14}},\ \bibinfo {pages}
  {4977} (\bibinfo {year} {2014})}\BibitemShut {NoStop}%
\bibitem [{\citenamefont {Önder Gül}\ \emph {et~al.}(2015)\citenamefont
  {Önder Gül}, \citenamefont {van Woerkom}, \citenamefont {van Weperen},
  \citenamefont {Car}, \citenamefont {Plissard}, \citenamefont {Bakkers},\ and\
  \citenamefont {Kouwenhoven}}]{Guel2015}%
  \BibitemOpen
  \bibfield  {author} {\bibinfo {author} {\bibnamefont {Önder Gül}}, \bibinfo
  {author} {\bibfnamefont {D.~J.}\ \bibnamefont {van Woerkom}}, \bibinfo
  {author} {\bibfnamefont {I.}~\bibnamefont {van Weperen}}, \bibinfo {author}
  {\bibfnamefont {D.}~\bibnamefont {Car}}, \bibinfo {author} {\bibfnamefont
  {S.~R.}\ \bibnamefont {Plissard}}, \bibinfo {author} {\bibfnamefont {E.~P.
  A.~M.}\ \bibnamefont {Bakkers}},\ and\ \bibinfo {author} {\bibfnamefont
  {L.~P.}\ \bibnamefont {Kouwenhoven}},\ }\bibfield  {title} {\bibinfo {title}
  {{Towards high mobility {InSb} nanowire devices}},\ }\href
  {https://doi.org/10.1088/0957-4484/26/21/215202} {\bibfield  {journal}
  {\bibinfo  {journal} {Nanotechnology}\ }\textbf {\bibinfo {volume} {26}},\
  \bibinfo {pages} {215202} (\bibinfo {year} {2015})}\BibitemShut {NoStop}%
\bibitem [{\citenamefont {Schroer}\ and\ \citenamefont
  {Petta}(2010)}]{Schroer2010}%
  \BibitemOpen
  \bibfield  {author} {\bibinfo {author} {\bibfnamefont {M.~D.}\ \bibnamefont
  {Schroer}}\ and\ \bibinfo {author} {\bibfnamefont {J.~R.}\ \bibnamefont
  {Petta}},\ }\bibfield  {title} {\bibinfo {title} {{Correlating the
  Nanostructure and Electronic Properties of InAs Nanowires}},\ }\href
  {https://doi.org/10.1021/nl904053j} {\bibfield  {journal} {\bibinfo
  {journal} {Nano Lett.}\ }\textbf {\bibinfo {volume} {10}},\ \bibinfo {pages}
  {1618} (\bibinfo {year} {2010})}\BibitemShut {NoStop}%
\bibitem [{\citenamefont {Wang}\ \emph {et~al.}(2013)\citenamefont {Wang},
  \citenamefont {Yip}, \citenamefont {Han}, \citenamefont {Fok}, \citenamefont
  {Lin}, \citenamefont {Hou}, \citenamefont {Dong}, \citenamefont {Hung},
  \citenamefont {Chan},\ and\ \citenamefont {Ho}}]{Wang2013}%
  \BibitemOpen
  \bibfield  {author} {\bibinfo {author} {\bibfnamefont {F.}~\bibnamefont
  {Wang}}, \bibinfo {author} {\bibfnamefont {S.}~\bibnamefont {Yip}}, \bibinfo
  {author} {\bibfnamefont {N.}~\bibnamefont {Han}}, \bibinfo {author}
  {\bibfnamefont {K.}~\bibnamefont {Fok}}, \bibinfo {author} {\bibfnamefont
  {H.}~\bibnamefont {Lin}}, \bibinfo {author} {\bibfnamefont {J.~J.}\
  \bibnamefont {Hou}}, \bibinfo {author} {\bibfnamefont {G.}~\bibnamefont
  {Dong}}, \bibinfo {author} {\bibfnamefont {T.}~\bibnamefont {Hung}}, \bibinfo
  {author} {\bibfnamefont {K.~S.}\ \bibnamefont {Chan}},\ and\ \bibinfo
  {author} {\bibfnamefont {J.~C.}\ \bibnamefont {Ho}},\ }\bibfield  {title}
  {\bibinfo {title} {{Surface roughness induced electron mobility degradation
  in {InAs} nanowires}},\ }\href
  {https://doi.org/10.1088/0957-4484/24/37/375202} {\bibfield  {journal}
  {\bibinfo  {journal} {Nanotechnology}\ }\textbf {\bibinfo {volume} {24}},\
  \bibinfo {pages} {375202} (\bibinfo {year} {2013})}\BibitemShut {NoStop}%
\bibitem [{\citenamefont {Gupta}\ \emph {et~al.}(2013)\citenamefont {Gupta},
  \citenamefont {Song}, \citenamefont {Holloway}, \citenamefont {Sinha},
  \citenamefont {Haapamaki}, \citenamefont {LaPierre},\ and\ \citenamefont
  {Baugh}}]{Gupta2013}%
  \BibitemOpen
  \bibfield  {author} {\bibinfo {author} {\bibfnamefont {N.}~\bibnamefont
  {Gupta}}, \bibinfo {author} {\bibfnamefont {Y.}~\bibnamefont {Song}},
  \bibinfo {author} {\bibfnamefont {G.~W.}\ \bibnamefont {Holloway}}, \bibinfo
  {author} {\bibfnamefont {U.}~\bibnamefont {Sinha}}, \bibinfo {author}
  {\bibfnamefont {C.~M.}\ \bibnamefont {Haapamaki}}, \bibinfo {author}
  {\bibfnamefont {R.~R.}\ \bibnamefont {LaPierre}},\ and\ \bibinfo {author}
  {\bibfnamefont {J.}~\bibnamefont {Baugh}},\ }\bibfield  {title} {\bibinfo
  {title} {Temperature-dependent electron mobility in {InAs} nanowires},\
  }\href {https://doi.org/10.1088/0957-4484/24/22/225202} {\bibfield  {journal}
  {\bibinfo  {journal} {Nanotechnology}\ }\textbf {\bibinfo {volume} {24}},\
  \bibinfo {pages} {225202} (\bibinfo {year} {2013})}\BibitemShut {NoStop}%
\bibitem [{\citenamefont {Sourribes}\ \emph {et~al.}(2014)\citenamefont
  {Sourribes}, \citenamefont {Isakov}, \citenamefont {Panfilova}, \citenamefont
  {Liu},\ and\ \citenamefont {Warburton}}]{Sourribes2014}%
  \BibitemOpen
  \bibfield  {author} {\bibinfo {author} {\bibfnamefont {M.~J.~L.}\
  \bibnamefont {Sourribes}}, \bibinfo {author} {\bibfnamefont {I.}~\bibnamefont
  {Isakov}}, \bibinfo {author} {\bibfnamefont {M.}~\bibnamefont {Panfilova}},
  \bibinfo {author} {\bibfnamefont {H.}~\bibnamefont {Liu}},\ and\ \bibinfo
  {author} {\bibfnamefont {P.~A.}\ \bibnamefont {Warburton}},\ }\bibfield
  {title} {\bibinfo {title} {{Mobility Enhancement by Sb-mediated Minimisation
  of Stacking Fault Density in InAs Nanowires Grown on Silicon}},\ }\href
  {https://doi.org/10.1021/nl5001554} {\bibfield  {journal} {\bibinfo
  {journal} {Nano Lett.}\ }\textbf {\bibinfo {volume} {14}},\ \bibinfo {pages}
  {1643} (\bibinfo {year} {2014})}\BibitemShut {NoStop}%
\bibitem [{\citenamefont {Blonder}\ \emph {et~al.}(1982)\citenamefont
  {Blonder}, \citenamefont {Tinkham},\ and\ \citenamefont
  {Klapwijk}}]{Blonder1982}%
  \BibitemOpen
  \bibfield  {author} {\bibinfo {author} {\bibfnamefont {G.~E.}\ \bibnamefont
  {Blonder}}, \bibinfo {author} {\bibfnamefont {M.}~\bibnamefont {Tinkham}},\
  and\ \bibinfo {author} {\bibfnamefont {T.~M.}\ \bibnamefont {Klapwijk}},\
  }\bibfield  {title} {\bibinfo {title} {{Transition from metallic to tunneling
  regimes in superconducting microconstrictions: Excess current, charge
  imbalance, and supercurrent conversion}},\ }\href
  {https://doi.org/10.1103/PhysRevB.25.4515} {\bibfield  {journal} {\bibinfo
  {journal} {Phys. Rev. B}\ }\textbf {\bibinfo {volume} {25}},\ \bibinfo
  {pages} {4515} (\bibinfo {year} {1982})}\BibitemShut {NoStop}%
\bibitem [{\citenamefont {Beenakker}(1992)}]{Beenaker1992}%
  \BibitemOpen
  \bibfield  {author} {\bibinfo {author} {\bibfnamefont {C.~W.~J.}\
  \bibnamefont {Beenakker}},\ }\bibfield  {title} {\bibinfo {title} {{Quantum
  transport in semiconductor-superconductor microjunctions}},\ }\href
  {https://doi.org/10.1103/PhysRevB.46.12841} {\bibfield  {journal} {\bibinfo
  {journal} {Phys. Rev. B}\ }\textbf {\bibinfo {volume} {46}},\ \bibinfo
  {pages} {12841} (\bibinfo {year} {1992})}\BibitemShut {NoStop}%
\bibitem [{\citenamefont {Octavio}\ \emph {et~al.}(1983)\citenamefont
  {Octavio}, \citenamefont {Tinkham}, \citenamefont {Blonder},\ and\
  \citenamefont {Klapwijk}}]{Octavio1983}%
  \BibitemOpen
  \bibfield  {author} {\bibinfo {author} {\bibfnamefont {M.}~\bibnamefont
  {Octavio}}, \bibinfo {author} {\bibfnamefont {M.}~\bibnamefont {Tinkham}},
  \bibinfo {author} {\bibfnamefont {G.~E.}\ \bibnamefont {Blonder}},\ and\
  \bibinfo {author} {\bibfnamefont {T.~M.}\ \bibnamefont {Klapwijk}},\
  }\bibfield  {title} {\bibinfo {title} {{Subharmonic energy-gap structure in
  superconducting constrictions}},\ }\href
  {https://doi.org/10.1103/PhysRevB.27.6739} {\bibfield  {journal} {\bibinfo
  {journal} {Phys. Rev. B}\ }\textbf {\bibinfo {volume} {27}},\ \bibinfo
  {pages} {6739} (\bibinfo {year} {1983})}\BibitemShut {NoStop}%
\bibitem [{\citenamefont {Aminov}\ \emph {et~al.}(1996)\citenamefont {Aminov},
  \citenamefont {Golubov},\ and\ \citenamefont {Kupriyanov}}]{Goloubov1996}%
  \BibitemOpen
  \bibfield  {author} {\bibinfo {author} {\bibfnamefont {B.~A.}\ \bibnamefont
  {Aminov}}, \bibinfo {author} {\bibfnamefont {A.~A.}\ \bibnamefont
  {Golubov}},\ and\ \bibinfo {author} {\bibfnamefont {M.~Y.}\ \bibnamefont
  {Kupriyanov}},\ }\bibfield  {title} {\bibinfo {title} {{Quasiparticle current
  in ballistic constrictions with finite transparencies of interfaces}},\
  }\href {https://doi.org/10.1103/PhysRevB.53.365} {\bibfield  {journal}
  {\bibinfo  {journal} {Phys. Rev. B}\ }\textbf {\bibinfo {volume} {53}},\
  \bibinfo {pages} {365} (\bibinfo {year} {1996})}\BibitemShut {NoStop}%
\bibitem [{\citenamefont {Tinkham}(2004)}]{Tinkham2004}%
  \BibitemOpen
  \bibfield  {author} {\bibinfo {author} {\bibfnamefont {M.}~\bibnamefont
  {Tinkham}},\ }\href {http://www.worldcat.org/isbn/0486435032} {\emph
  {\bibinfo {title} {{Introduction to Superconductivity}}}},\ \bibinfo
  {edition} {2nd}\ ed.\ (\bibinfo  {publisher} {Dover Publications},\ \bibinfo
  {year} {2004})\BibitemShut {NoStop}%
\bibitem [{\citenamefont {Jarillo-Herrero}\ \emph {et~al.}(2006)\citenamefont
  {Jarillo-Herrero}, \citenamefont {van Dam},\ and\ \citenamefont
  {Kouwenhoven}}]{Jarillo-Herrero2006}%
  \BibitemOpen
  \bibfield  {author} {\bibinfo {author} {\bibfnamefont {P.}~\bibnamefont
  {Jarillo-Herrero}}, \bibinfo {author} {\bibfnamefont {J.~A.}\ \bibnamefont
  {van Dam}},\ and\ \bibinfo {author} {\bibfnamefont {L.~P.}\ \bibnamefont
  {Kouwenhoven}},\ }\bibfield  {title} {\bibinfo {title} {{Quantum supercurrent
  transistors in carbon nanotubes}},\ }\href
  {https://doi.org/10.1038/nature04550} {\bibfield  {journal} {\bibinfo
  {journal} {Nature}\ }\textbf {\bibinfo {volume} {439}},\ \bibinfo {pages}
  {953} (\bibinfo {year} {2006})}\BibitemShut {NoStop}%
\bibitem [{\citenamefont {Flensberg}\ \emph {et~al.}(1988)\citenamefont
  {Flensberg}, \citenamefont {Hansen},\ and\ \citenamefont
  {Octavio}}]{Flensberg1988}%
  \BibitemOpen
  \bibfield  {author} {\bibinfo {author} {\bibfnamefont {K.}~\bibnamefont
  {Flensberg}}, \bibinfo {author} {\bibfnamefont {J.~B.}\ \bibnamefont
  {Hansen}},\ and\ \bibinfo {author} {\bibfnamefont {M.}~\bibnamefont
  {Octavio}},\ }\bibfield  {title} {\bibinfo {title} {{Subharmonic energy-gap
  structure in superconducting weak links}},\ }\href
  {https://doi.org/10.1103/PhysRevB.38.8707} {\bibfield  {journal} {\bibinfo
  {journal} {Phys. Rev. B}\ }\textbf {\bibinfo {volume} {38}},\ \bibinfo
  {pages} {8707} (\bibinfo {year} {1988})}\BibitemShut {NoStop}%
\bibitem [{\citenamefont {Eiles}\ and\ \citenamefont
  {Martinis}(1994)}]{Eiles1994}%
  \BibitemOpen
  \bibfield  {author} {\bibinfo {author} {\bibfnamefont {T.~M.}\ \bibnamefont
  {Eiles}}\ and\ \bibinfo {author} {\bibfnamefont {J.~M.}\ \bibnamefont
  {Martinis}},\ }\bibfield  {title} {\bibinfo {title} {{Combined Josephson and
  charging behavior of the supercurrent in the superconducting single-electron
  transistor}},\ }\href {https://doi.org/10.1103/PhysRevB.50.627} {\bibfield
  {journal} {\bibinfo  {journal} {Phys. Rev. B}\ }\textbf {\bibinfo {volume}
  {50}},\ \bibinfo {pages} {627} (\bibinfo {year} {1994})}\BibitemShut
  {NoStop}%
\bibitem [{\citenamefont {Ridderbos}\ \emph {et~al.}(2018)\citenamefont
  {Ridderbos}, \citenamefont {Brauns}, \citenamefont {Shen}, \citenamefont
  {de~Vries}, \citenamefont {Li}, \citenamefont {Bakkers}, \citenamefont
  {Brinkman},\ and\ \citenamefont {Zwanenburg}}]{Ridderbos2018}%
  \BibitemOpen
  \bibfield  {author} {\bibinfo {author} {\bibfnamefont {J.}~\bibnamefont
  {Ridderbos}}, \bibinfo {author} {\bibfnamefont {M.}~\bibnamefont {Brauns}},
  \bibinfo {author} {\bibfnamefont {J.}~\bibnamefont {Shen}}, \bibinfo {author}
  {\bibfnamefont {F.~K.}\ \bibnamefont {de~Vries}}, \bibinfo {author}
  {\bibfnamefont {A.}~\bibnamefont {Li}}, \bibinfo {author} {\bibfnamefont
  {E.~P. A.~M.}\ \bibnamefont {Bakkers}}, \bibinfo {author} {\bibfnamefont
  {A.}~\bibnamefont {Brinkman}},\ and\ \bibinfo {author} {\bibfnamefont
  {F.~A.}\ \bibnamefont {Zwanenburg}},\ }\bibfield  {title} {\bibinfo {title}
  {{Josephson Effect in a Few-Hole Quantum Dot}},\ }\href
  {https://doi.org/10.1002/adma.201802257} {\bibfield  {journal} {\bibinfo
  {journal} {Adv. Mater.}\ }\textbf {\bibinfo {volume} {30}},\ \bibinfo {pages}
  {1802257} (\bibinfo {year} {2018})}\BibitemShut {NoStop}%
\bibitem [{\citenamefont {Tafuri}(2019)}]{Tafuri2019}%
  \BibitemOpen
  \bibfield  {author} {\bibinfo {author} {\bibfnamefont {F.}~\bibnamefont
  {Tafuri}},\ }\href
  {https://www.ebook.de/de/product/36577666/fundamentals_and_frontiers_of_the_josephson_effect.html}
  {\emph {\bibinfo {title} {{Fundamentals and Frontiers of the Josephson
  Effect}}}}\ (\bibinfo  {publisher} {Springer International Publishing},\
  \bibinfo {year} {2019})\BibitemShut {NoStop}%
\bibitem [{\citenamefont {Abay}\ \emph {et~al.}(2014)\citenamefont {Abay},
  \citenamefont {Persson}, \citenamefont {Nilsson}, \citenamefont {Wu},
  \citenamefont {Xu}, \citenamefont {Fogelstr\"om}, \citenamefont {Shumeiko},\
  and\ \citenamefont {Delsing}}]{Abay2014}%
  \BibitemOpen
  \bibfield  {author} {\bibinfo {author} {\bibfnamefont {S.}~\bibnamefont
  {Abay}}, \bibinfo {author} {\bibfnamefont {D.}~\bibnamefont {Persson}},
  \bibinfo {author} {\bibfnamefont {H.}~\bibnamefont {Nilsson}}, \bibinfo
  {author} {\bibfnamefont {F.}~\bibnamefont {Wu}}, \bibinfo {author}
  {\bibfnamefont {H.~Q.}\ \bibnamefont {Xu}}, \bibinfo {author} {\bibfnamefont
  {M.}~\bibnamefont {Fogelstr\"om}}, \bibinfo {author} {\bibfnamefont
  {V.}~\bibnamefont {Shumeiko}},\ and\ \bibinfo {author} {\bibfnamefont
  {P.}~\bibnamefont {Delsing}},\ }\bibfield  {title} {\bibinfo {title} {{Charge
  transport in InAs nanowire Josephson junctions}},\ }\href
  {https://doi.org/10.1103/PhysRevB.89.214508} {\bibfield  {journal} {\bibinfo
  {journal} {Phys. Rev. B}\ }\textbf {\bibinfo {volume} {89}},\ \bibinfo
  {pages} {214508} (\bibinfo {year} {2014})}\BibitemShut {NoStop}%
\bibitem [{\citenamefont {Doh}\ \emph {et~al.}(2005)\citenamefont {Doh},
  \citenamefont {van Dam}, \citenamefont {Roest}, \citenamefont {Bakkers},
  \citenamefont {Kouwenhoven},\ and\ \citenamefont {De~Franceschi}}]{Doh2005}%
  \BibitemOpen
  \bibfield  {author} {\bibinfo {author} {\bibfnamefont {Y.-J.}\ \bibnamefont
  {Doh}}, \bibinfo {author} {\bibfnamefont {J.~A.}\ \bibnamefont {van Dam}},
  \bibinfo {author} {\bibfnamefont {A.~L.}\ \bibnamefont {Roest}}, \bibinfo
  {author} {\bibfnamefont {E.~P. A.~M.}\ \bibnamefont {Bakkers}}, \bibinfo
  {author} {\bibfnamefont {L.~P.}\ \bibnamefont {Kouwenhoven}},\ and\ \bibinfo
  {author} {\bibfnamefont {S.}~\bibnamefont {De~Franceschi}},\ }\bibfield
  {title} {\bibinfo {title} {{Tunable Supercurrent Through Semiconductor
  Nanowires}},\ }\href {https://doi.org/10.1126/science.1113523} {\bibfield
  {journal} {\bibinfo  {journal} {Science}\ }\textbf {\bibinfo {volume}
  {309}},\ \bibinfo {pages} {272} (\bibinfo {year} {2005})}\BibitemShut
  {NoStop}%
\bibitem [{\citenamefont {Montemurro}\ \emph {et~al.}(2015)\citenamefont
  {Montemurro}, \citenamefont {Stornaiuolo}, \citenamefont {Massarotti},
  \citenamefont {Ercolani}, \citenamefont {Sorba}, \citenamefont {Beltram},
  \citenamefont {Tafuri},\ and\ \citenamefont {Roddaro}}]{Montemurro2015}%
  \BibitemOpen
  \bibfield  {author} {\bibinfo {author} {\bibfnamefont {D.}~\bibnamefont
  {Montemurro}}, \bibinfo {author} {\bibfnamefont {D.}~\bibnamefont
  {Stornaiuolo}}, \bibinfo {author} {\bibfnamefont {D.}~\bibnamefont
  {Massarotti}}, \bibinfo {author} {\bibfnamefont {D.}~\bibnamefont
  {Ercolani}}, \bibinfo {author} {\bibfnamefont {L.}~\bibnamefont {Sorba}},
  \bibinfo {author} {\bibfnamefont {F.}~\bibnamefont {Beltram}}, \bibinfo
  {author} {\bibfnamefont {F.}~\bibnamefont {Tafuri}},\ and\ \bibinfo {author}
  {\bibfnamefont {S.}~\bibnamefont {Roddaro}},\ }\bibfield  {title} {\bibinfo
  {title} {{Suspended {InAs} nanowire Josephson junctions assembled via
  dielectrophoresis}},\ }\href {https://doi.org/10.1088/0957-4484/26/38/385302}
  {\bibfield  {journal} {\bibinfo  {journal} {Nanotechnology}\ }\textbf
  {\bibinfo {volume} {26}},\ \bibinfo {pages} {385302} (\bibinfo {year}
  {2015})}\BibitemShut {NoStop}%
\bibitem [{\citenamefont {Frielinghaus}\ \emph {et~al.}(2010)\citenamefont
  {Frielinghaus}, \citenamefont {Batov}, \citenamefont {Weides}, \citenamefont
  {Kohlstedt}, \citenamefont {Calarco},\ and\ \citenamefont
  {Schäpers}}]{Frielinghaus2010}%
  \BibitemOpen
  \bibfield  {author} {\bibinfo {author} {\bibfnamefont {R.}~\bibnamefont
  {Frielinghaus}}, \bibinfo {author} {\bibfnamefont {I.~E.}\ \bibnamefont
  {Batov}}, \bibinfo {author} {\bibfnamefont {M.}~\bibnamefont {Weides}},
  \bibinfo {author} {\bibfnamefont {H.}~\bibnamefont {Kohlstedt}}, \bibinfo
  {author} {\bibfnamefont {R.}~\bibnamefont {Calarco}},\ and\ \bibinfo {author}
  {\bibfnamefont {T.}~\bibnamefont {Schäpers}},\ }\bibfield  {title} {\bibinfo
  {title} {{Josephson supercurrent in Nb/{InN}-nanowire/Nb junctions}},\ }\href
  {https://doi.org/10.1063/1.3377897} {\bibfield  {journal} {\bibinfo
  {journal} {Appl. Phys. Lett.}\ }\textbf {\bibinfo {volume} {96}},\ \bibinfo
  {pages} {132504} (\bibinfo {year} {2010})}\BibitemShut {NoStop}%
\bibitem [{\citenamefont {Ojeda-Aristizabal}\ \emph {et~al.}(2009)\citenamefont
  {Ojeda-Aristizabal}, \citenamefont {Ferrier}, \citenamefont {Gu\'eron},\ and\
  \citenamefont {Bouchiat}}]{Ojeda-Aristizabal2009}%
  \BibitemOpen
  \bibfield  {author} {\bibinfo {author} {\bibfnamefont {C.}~\bibnamefont
  {Ojeda-Aristizabal}}, \bibinfo {author} {\bibfnamefont {M.}~\bibnamefont
  {Ferrier}}, \bibinfo {author} {\bibfnamefont {S.}~\bibnamefont {Gu\'eron}},\
  and\ \bibinfo {author} {\bibfnamefont {H.}~\bibnamefont {Bouchiat}},\
  }\bibfield  {title} {\bibinfo {title} {{Tuning the proximity effect in a
  superconductor-graphene-superconductor junction}},\ }\href
  {https://doi.org/10.1103/PhysRevB.79.165436} {\bibfield  {journal} {\bibinfo
  {journal} {Phys. Rev. B}\ }\textbf {\bibinfo {volume} {79}},\ \bibinfo
  {pages} {165436} (\bibinfo {year} {2009})}\BibitemShut {NoStop}%
\bibitem [{\citenamefont {Christian~Enss}(2005)}]{Enss2005}%
  \BibitemOpen
  \bibfield  {author} {\bibinfo {author} {\bibfnamefont {S.~H.}\ \bibnamefont
  {Christian~Enss}},\ }\href
  {https://www.ebook.de/de/product/3270653/christian_enss_siegfried_hunklinger_low_temperature_physics.html}
  {\emph {\bibinfo {title} {{Low-Temperature Physics}}}}\ (\bibinfo
  {publisher} {Springer Berlin Heidelberg},\ \bibinfo {year}
  {2005})\BibitemShut {NoStop}%
\bibitem [{\citenamefont {Kupriyanov}\ \emph {et~al.}(1999)\citenamefont
  {Kupriyanov}, \citenamefont {Brinkman}, \citenamefont {Golubov},
  \citenamefont {Siegel},\ and\ \citenamefont {Rogalla}}]{Kupriyanov1999}%
  \BibitemOpen
  \bibfield  {author} {\bibinfo {author} {\bibfnamefont {M.}~\bibnamefont
  {Kupriyanov}}, \bibinfo {author} {\bibfnamefont {A.}~\bibnamefont
  {Brinkman}}, \bibinfo {author} {\bibfnamefont {A.}~\bibnamefont {Golubov}},
  \bibinfo {author} {\bibfnamefont {M.}~\bibnamefont {Siegel}},\ and\ \bibinfo
  {author} {\bibfnamefont {H.}~\bibnamefont {Rogalla}},\ }\bibfield  {title}
  {\bibinfo {title} {{Double-barrier Josephson structures as the novel elements
  for superconducting large-scale integrated circuits}},\ }\href
  {https://doi.org/10.1016/s0921-4534(99)00408-6} {\bibfield  {journal}
  {\bibinfo  {journal} {Physica C Supercond.}\ }\textbf {\bibinfo {volume}
  {326-327}},\ \bibinfo {pages} {16} (\bibinfo {year} {1999})}\BibitemShut
  {NoStop}%
\bibitem [{\citenamefont {Dubos}\ \emph {et~al.}(2001)\citenamefont {Dubos},
  \citenamefont {Courtois}, \citenamefont {Pannetier}, \citenamefont {Wilhelm},
  \citenamefont {Zaikin},\ and\ \citenamefont {Sch\"on}}]{Dubos2001}%
  \BibitemOpen
  \bibfield  {author} {\bibinfo {author} {\bibfnamefont {P.}~\bibnamefont
  {Dubos}}, \bibinfo {author} {\bibfnamefont {H.}~\bibnamefont {Courtois}},
  \bibinfo {author} {\bibfnamefont {B.}~\bibnamefont {Pannetier}}, \bibinfo
  {author} {\bibfnamefont {F.~K.}\ \bibnamefont {Wilhelm}}, \bibinfo {author}
  {\bibfnamefont {A.~D.}\ \bibnamefont {Zaikin}},\ and\ \bibinfo {author}
  {\bibfnamefont {G.}~\bibnamefont {Sch\"on}},\ }\bibfield  {title} {\bibinfo
  {title} {{Josephson critical current in a long mesoscopic S-N-S junction}},\
  }\href {https://doi.org/10.1103/PhysRevB.63.064502} {\bibfield  {journal}
  {\bibinfo  {journal} {Phys. Rev. B}\ }\textbf {\bibinfo {volume} {63}},\
  \bibinfo {pages} {064502} (\bibinfo {year} {2001})}\BibitemShut {NoStop}%
\bibitem [{\citenamefont {Mayer}\ \emph {et~al.}(2019)\citenamefont {Mayer},
  \citenamefont {Yuan}, \citenamefont {Wickramasinghe}, \citenamefont {Nguyen},
  \citenamefont {Dartiailh},\ and\ \citenamefont {Shabani}}]{Mayer2019}%
  \BibitemOpen
  \bibfield  {author} {\bibinfo {author} {\bibfnamefont {W.}~\bibnamefont
  {Mayer}}, \bibinfo {author} {\bibfnamefont {J.}~\bibnamefont {Yuan}},
  \bibinfo {author} {\bibfnamefont {K.~S.}\ \bibnamefont {Wickramasinghe}},
  \bibinfo {author} {\bibfnamefont {T.}~\bibnamefont {Nguyen}}, \bibinfo
  {author} {\bibfnamefont {M.~C.}\ \bibnamefont {Dartiailh}},\ and\ \bibinfo
  {author} {\bibfnamefont {J.}~\bibnamefont {Shabani}},\ }\bibfield  {title}
  {\bibinfo {title} {{Superconducting proximity effect in epitaxial Al-{InAs}
  heterostructures}},\ }\href {https://doi.org/10.1063/1.5067363} {\bibfield
  {journal} {\bibinfo  {journal} {Appl. Phys. Lett.}\ }\textbf {\bibinfo
  {volume} {114}},\ \bibinfo {pages} {103104} (\bibinfo {year}
  {2019})}\BibitemShut {NoStop}%
\bibitem [{\citenamefont {Averin}\ and\ \citenamefont
  {Bardas}(1995)}]{Averin1995}%
  \BibitemOpen
  \bibfield  {author} {\bibinfo {author} {\bibfnamefont {D.}~\bibnamefont
  {Averin}}\ and\ \bibinfo {author} {\bibfnamefont {A.}~\bibnamefont
  {Bardas}},\ }\bibfield  {title} {\bibinfo {title} {{ac Josephson Effect in a
  Single Quantum Channel}},\ }\href
  {https://doi.org/10.1103/PhysRevLett.75.1831} {\bibfield  {journal} {\bibinfo
   {journal} {Phys. Rev. Lett.}\ }\textbf {\bibinfo {volume} {75}},\ \bibinfo
  {pages} {1831} (\bibinfo {year} {1995})}\BibitemShut {NoStop}%
\bibitem [{\citenamefont {Kjaergaard}\ \emph {et~al.}(2017)\citenamefont
  {Kjaergaard}, \citenamefont {Suominen}, \citenamefont {Nowak}, \citenamefont
  {Akhmerov}, \citenamefont {Shabani}, \citenamefont {Palmstr\o{}m},
  \citenamefont {Nichele},\ and\ \citenamefont {Marcus}}]{Kjaergaard2017}%
  \BibitemOpen
  \bibfield  {author} {\bibinfo {author} {\bibfnamefont {M.}~\bibnamefont
  {Kjaergaard}}, \bibinfo {author} {\bibfnamefont {H.~J.}\ \bibnamefont
  {Suominen}}, \bibinfo {author} {\bibfnamefont {M.~P.}\ \bibnamefont {Nowak}},
  \bibinfo {author} {\bibfnamefont {A.~R.}\ \bibnamefont {Akhmerov}}, \bibinfo
  {author} {\bibfnamefont {J.}~\bibnamefont {Shabani}}, \bibinfo {author}
  {\bibfnamefont {C.~J.}\ \bibnamefont {Palmstr\o{}m}}, \bibinfo {author}
  {\bibfnamefont {F.}~\bibnamefont {Nichele}},\ and\ \bibinfo {author}
  {\bibfnamefont {C.~M.}\ \bibnamefont {Marcus}},\ }\bibfield  {title}
  {\bibinfo {title} {{Transparent Semiconductor-Superconductor Interface and
  Induced Gap in an Epitaxial Heterostructure Josephson Junction}},\ }\href
  {https://doi.org/10.1103/PhysRevApplied.7.034029} {\bibfield  {journal}
  {\bibinfo  {journal} {Phys. Rev. Appl.}\ }\textbf {\bibinfo {volume} {7}},\
  \bibinfo {pages} {034029} (\bibinfo {year} {2017})}\BibitemShut {NoStop}%
\bibitem [{\citenamefont {Hendrickx}\ \emph {et~al.}(2019)\citenamefont
  {Hendrickx}, \citenamefont {Tagliaferri}, \citenamefont {Kouwenhoven},
  \citenamefont {Li}, \citenamefont {Franke}, \citenamefont {Sammak},
  \citenamefont {Brinkman}, \citenamefont {Scappucci},\ and\ \citenamefont
  {Veldhorst}}]{Hendrickx2019}%
  \BibitemOpen
  \bibfield  {author} {\bibinfo {author} {\bibfnamefont {N.~W.}\ \bibnamefont
  {Hendrickx}}, \bibinfo {author} {\bibfnamefont {M.~L.~V.}\ \bibnamefont
  {Tagliaferri}}, \bibinfo {author} {\bibfnamefont {M.}~\bibnamefont
  {Kouwenhoven}}, \bibinfo {author} {\bibfnamefont {R.}~\bibnamefont {Li}},
  \bibinfo {author} {\bibfnamefont {D.~P.}\ \bibnamefont {Franke}}, \bibinfo
  {author} {\bibfnamefont {A.}~\bibnamefont {Sammak}}, \bibinfo {author}
  {\bibfnamefont {A.}~\bibnamefont {Brinkman}}, \bibinfo {author}
  {\bibfnamefont {G.}~\bibnamefont {Scappucci}},\ and\ \bibinfo {author}
  {\bibfnamefont {M.}~\bibnamefont {Veldhorst}},\ }\bibfield  {title} {\bibinfo
  {title} {{Ballistic supercurrent discretization and micrometer-long Josephson
  coupling in germanium}},\ }\href {https://doi.org/10.1103/PhysRevB.99.075435}
  {\bibfield  {journal} {\bibinfo  {journal} {Phys. Rev. B}\ }\textbf {\bibinfo
  {volume} {99}},\ \bibinfo {pages} {075435} (\bibinfo {year}
  {2019})}\BibitemShut {NoStop}%
\bibitem [{\citenamefont {Ridderbos}\ \emph {et~al.}(2019)\citenamefont
  {Ridderbos}, \citenamefont {Brauns}, \citenamefont {Li}, \citenamefont
  {Bakkers}, \citenamefont {Brinkman}, \citenamefont {van~der Wiel},\ and\
  \citenamefont {Zwanenburg}}]{Ridderbos2019}%
  \BibitemOpen
  \bibfield  {author} {\bibinfo {author} {\bibfnamefont {J.}~\bibnamefont
  {Ridderbos}}, \bibinfo {author} {\bibfnamefont {M.}~\bibnamefont {Brauns}},
  \bibinfo {author} {\bibfnamefont {A.}~\bibnamefont {Li}}, \bibinfo {author}
  {\bibfnamefont {E.~P. A.~M.}\ \bibnamefont {Bakkers}}, \bibinfo {author}
  {\bibfnamefont {A.}~\bibnamefont {Brinkman}}, \bibinfo {author}
  {\bibfnamefont {W.~G.}\ \bibnamefont {van~der Wiel}},\ and\ \bibinfo {author}
  {\bibfnamefont {F.~A.}\ \bibnamefont {Zwanenburg}},\ }\bibfield  {title}
  {\bibinfo {title} {{Multiple Andreev reflections and Shapiro steps in a Ge-Si
  nanowire Josephson junction}},\ }\href
  {https://doi.org/10.1103/PhysRevMaterials.3.084803} {\bibfield  {journal}
  {\bibinfo  {journal} {Phys. Rev. Mater.}\ }\textbf {\bibinfo {volume} {3}},\
  \bibinfo {pages} {084803} (\bibinfo {year} {2019})}\BibitemShut {NoStop}%
\bibitem [{\citenamefont {Larsen}\ \emph {et~al.}(2020)\citenamefont {Larsen},
  \citenamefont {Gershenson}, \citenamefont {Casparis}, \citenamefont
  {Kringh\o{}j}, \citenamefont {Pearson}, \citenamefont {McNeil}, \citenamefont
  {Kuemmeth}, \citenamefont {Krogstrup}, \citenamefont {Petersson},\ and\
  \citenamefont {Marcus}}]{Larsen2020}%
  \BibitemOpen
  \bibfield  {author} {\bibinfo {author} {\bibfnamefont {T.~W.}\ \bibnamefont
  {Larsen}}, \bibinfo {author} {\bibfnamefont {M.~E.}\ \bibnamefont
  {Gershenson}}, \bibinfo {author} {\bibfnamefont {L.}~\bibnamefont
  {Casparis}}, \bibinfo {author} {\bibfnamefont {A.}~\bibnamefont
  {Kringh\o{}j}}, \bibinfo {author} {\bibfnamefont {N.~J.}\ \bibnamefont
  {Pearson}}, \bibinfo {author} {\bibfnamefont {R.~P.~G.}\ \bibnamefont
  {McNeil}}, \bibinfo {author} {\bibfnamefont {F.}~\bibnamefont {Kuemmeth}},
  \bibinfo {author} {\bibfnamefont {P.}~\bibnamefont {Krogstrup}}, \bibinfo
  {author} {\bibfnamefont {K.~D.}\ \bibnamefont {Petersson}},\ and\ \bibinfo
  {author} {\bibfnamefont {C.~M.}\ \bibnamefont {Marcus}},\ }\bibfield  {title}
  {\bibinfo {title} {Parity-protected superconductor-semiconductor qubit},\
  }\href {https://doi.org/10.1103/PhysRevLett.125.056801} {\bibfield  {journal}
  {\bibinfo  {journal} {Phys. Rev. Lett.}\ }\textbf {\bibinfo {volume} {125}},\
  \bibinfo {pages} {056801} (\bibinfo {year} {2020})}\BibitemShut {NoStop}%
\bibitem [{\citenamefont {Lutchyn}\ \emph {et~al.}(2018)\citenamefont
  {Lutchyn}, \citenamefont {Bakkers}, \citenamefont {Kouwenhoven},
  \citenamefont {Krogstrup}, \citenamefont {Marcus},\ and\ \citenamefont
  {Oreg}}]{Lutchyn2018}%
  \BibitemOpen
  \bibfield  {author} {\bibinfo {author} {\bibfnamefont {R.~M.}\ \bibnamefont
  {Lutchyn}}, \bibinfo {author} {\bibfnamefont {E.~P. A.~M.}\ \bibnamefont
  {Bakkers}}, \bibinfo {author} {\bibfnamefont {L.~P.}\ \bibnamefont
  {Kouwenhoven}}, \bibinfo {author} {\bibfnamefont {P.}~\bibnamefont
  {Krogstrup}}, \bibinfo {author} {\bibfnamefont {C.~M.}\ \bibnamefont
  {Marcus}},\ and\ \bibinfo {author} {\bibfnamefont {Y.}~\bibnamefont {Oreg}},\
  }\bibfield  {title} {\bibinfo {title} {Majorana zero modes in
  superconductor-semiconductor heterostructures},\ }\href
  {https://doi.org/10.1038/s41578-018-0003-1} {\bibfield  {journal} {\bibinfo
  {journal} {Nature Reviews Materials}\ }\textbf {\bibinfo {volume} {3}},\
  \bibinfo {pages} {52} (\bibinfo {year} {2018})}\BibitemShut {NoStop}%
\end{thebibliography}

%

\onecolumngrid
\clearpage
\onecolumngrid
\setcounter{figure}{0}
\setcounter{section}{0}
\setcounter{page}{1}
\renewcommand\thefigure{S\arabic{figure}}
\renewcommand{\thetable}{S\arabic{table}}
	
\section*{Supplemental Material for `Electrical Properties of Selective-Area-Grown Superconductor-Semiconductor Hybrid Structures on Silicon'}
	
\vspace{1cm}
	
\section{Gate capacitance simulations}\label{Gate_Capacitance_Appendix}

\begin{figure}[h!]
	\includegraphics{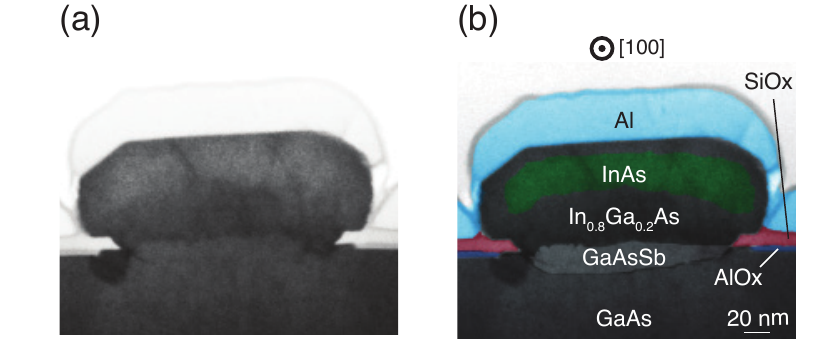}
	\caption{\label{fig:FigCapacitance} \textbf{(a)} Original and \textbf{(b)} false-colored scanning transmission electron micrograph of the material stack.} 
	\label{fig:false_color_comp}
\end{figure}

	Figure \ref{fig:false_color_comp}a shows a scanning transmission electron micrograph of a nanowire. Based on the contrast, the respective material layers are identified as shown in the false-colored image in Fig.~\ref{fig:false_color_comp}b.
	
\begin{figure}[h!]
	\includegraphics{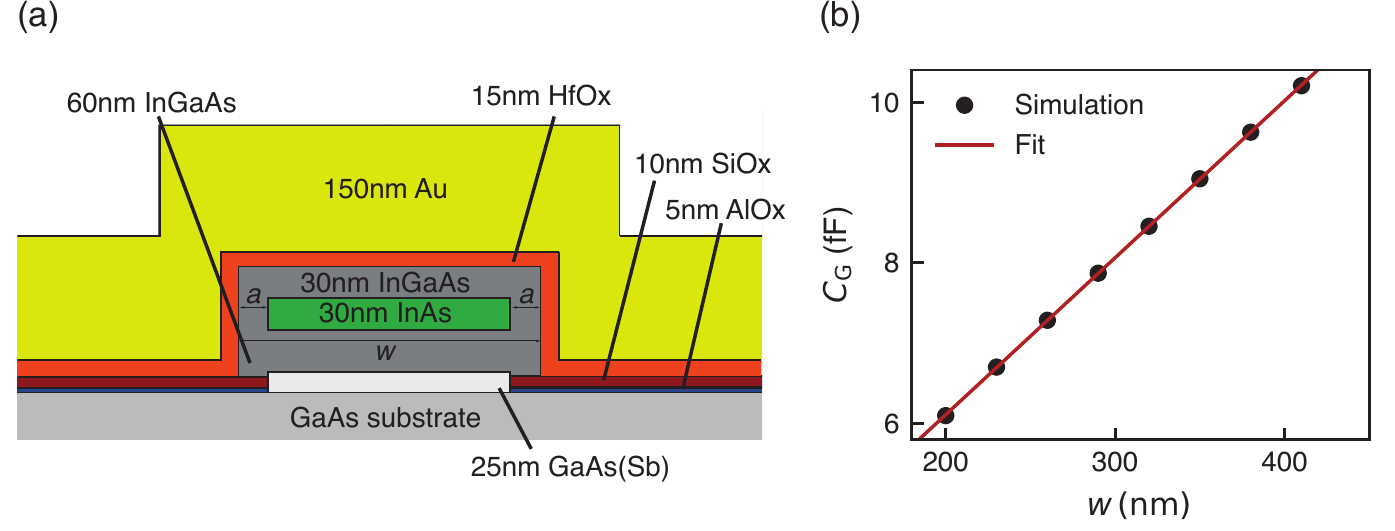}
	\caption{\label{fig:FigCapacitance} \textbf{(a)} Parameterized model used to simulate the capacitance between the conducting InAs layer and the Au topgate $C_{\mathrm{G}}$ for different wire widths $w$ using finite element methods, where $a=15\,\mathrm{nm}$ is assumed. \textbf{(b)} Simulation results for $C_{\mathrm{G}}$ as a function of $w$. A linear fit is added that gives $C_{\mathrm{G}} = w\cdot 0.06\,\mathrm{fF/nm} + 6.426\,\mathrm{fF}$ for $w>200\,\mathrm{nm}$.} 
	\label{fig:gatecapacitance}
\end{figure}

	To simulate the gate capacitance $C_{\mathrm{G}}$ between the conducting InAs layer and the topgate we used the finite element method in the electrostatic module of COMSOL Multiphysics. The cross-sectional geometry was based on the scanning transmission electron microscope images (Fig.~\ref{fig:false_color_comp}), where we assume that the InAs layer(green) is encapsulated in the InGaAs material stack (grey). This cross-section was approximated by a rectangular geometry as shown in Fig.~\ref{fig:gatecapacitance}a and extruded over a length of $L_{\mathrm{FET}} = 6\, \mathrm{\micro m}$. Based on the scanning transmission electron microscope images (Fig.~\ref{fig:false_color_comp}) we assume that the InAs layer (green) is encapsulated in the InGaAs material stack (grey). The gate capacitance $C_{\mathrm{G}}$ as a function of nanowire width $w$ is extracted from a linear fit to the simulation results of 8 different nanowire widths (Fig.~\ref{fig:gatecapacitance}b).
	
\section{Measurements and data extraction}

\begin{figure}[t!]
	\includegraphics[scale=0.8]{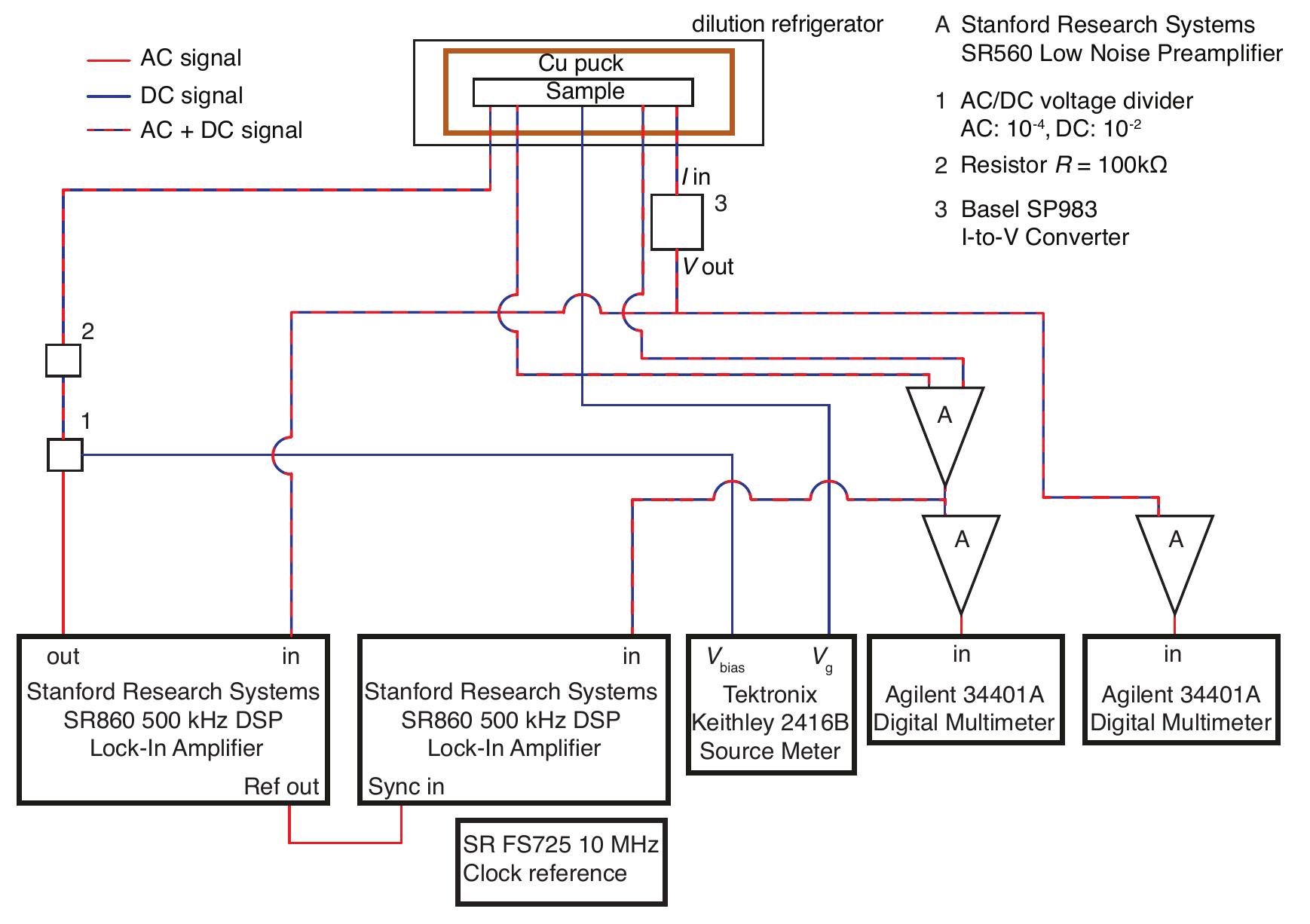}
	\caption{\label{fig:setup} Schematic of the instrumentation used for four-terminal, current biased, DC transport measurements. Red lines indicate cables that carry an AC signal, blue lines indicate lines that carry a DC signal and blue-red dashed lines indicate cables that carry both components. All instruments are synchronized with a $10\,\mathrm{MHz}$ clock reference. To measure the device in a 2-terminal voltage-bias configuration the bias resistor ($R = 100\,\mathrm{k\Omega}$) is removed and only the signal after the transimpedance amplifier (Basel SP983) is measured with the lock-in amplifier to obtain $\mathrm{d}I/\mathrm{d}V$.} 
\end{figure}

	All data in this work were taken in a dilution refrigerator with a base temperature of 20 mK. FET devices and NIS devices were measured in a 2-probe setup with standard lock-in techniques. The devices were voltage biased and the current measured with a transimpedance amplifier. All data were corrected for a constant line resistance $R_{\mathrm{line}} = 4.9\,\mathrm{k\Omega}$. During measurements of NIS devices and SSmS Josephson junctions the unused end of the device was set to a floating configuration to not affect the measurements. The SSmS JJ devices were measured in a four-probe setup using a current bias. The current bias was set by a high bias resistor in series with the device. While the current was measured using a transimpedance amplifier, the voltage across the device was measured with a differential amplifier. Here, the differential and absolute values of current and voltage were measured simultaneously. Figure~\ref{fig:setup} shows a schematic of the setup.
	
\subsection{Tunneling spectroscopy}
\begin{figure*}[h!]
	\includegraphics{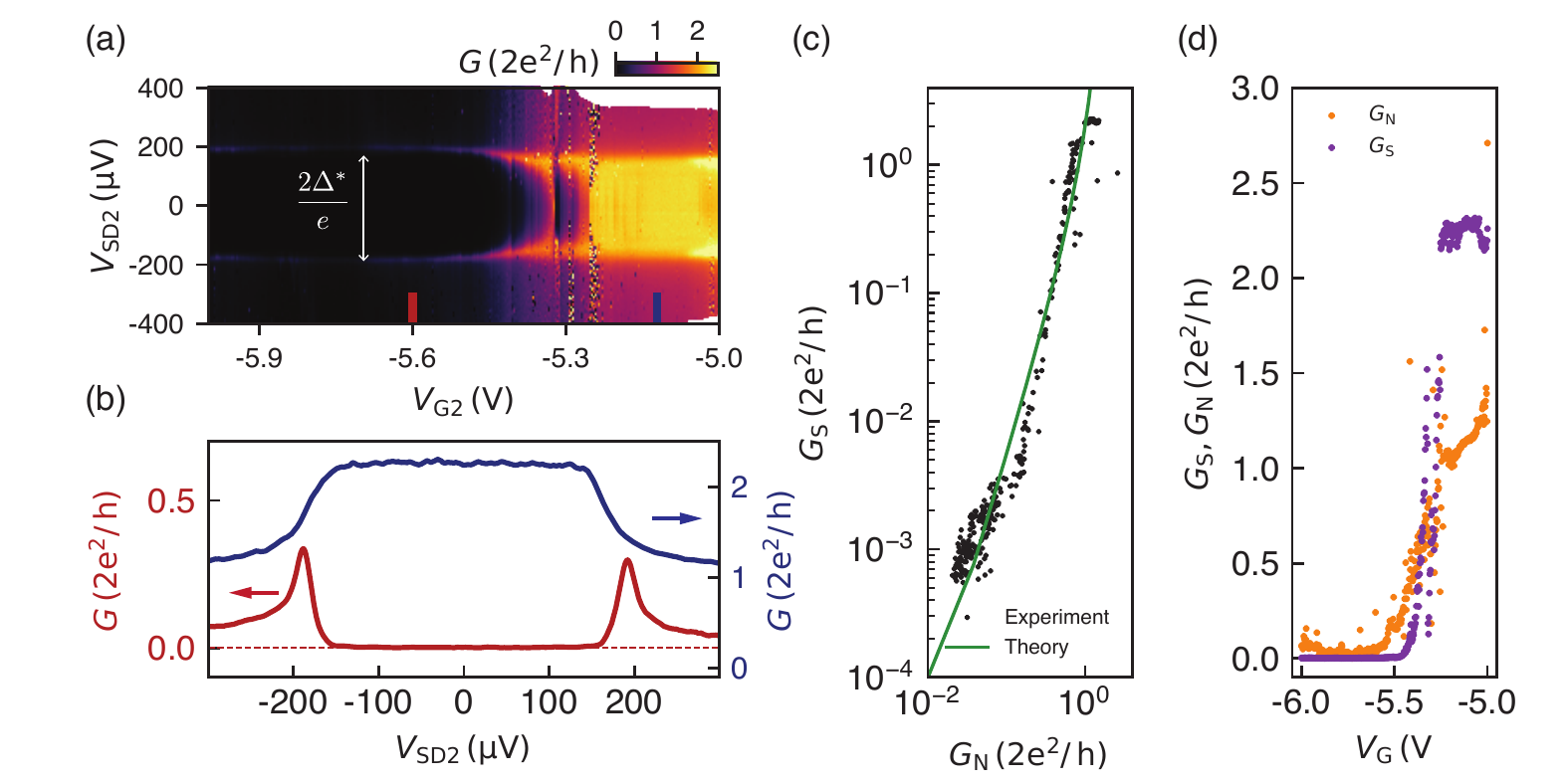}
	\caption{\label{fig:Fig5} \textbf{(a)} Differential conductance $\mathrm{d}I/\mathrm{d}V$ of device A as a function of gate voltage $V_\mathrm{G2}$ and source-drain voltage $V_{\mathrm{SD2}}$. Vertical cuts in the tunneling regime (red) and open regime (blue) are shown in \textbf{(b)}. To calculate $\mathrm{d}I/\mathrm{d}V$ the data were smoothed over 30 steps and the derivative of the current and voltage was calculated numerically. \textbf{{(c)}} Averaged differential conductance at zero source-drain voltage $G_{\mathrm{S}}$ versus averaged differential conductance at finite source-drain voltage $G_{\mathrm{N}}$ ($-350\,\mathrm{\micro V} < V_{\mathrm{SD2}} < -250\,\mathrm{\micro V}$). The green line is the theoretically predicted conductance in an Andreev enhanced QPC with no fitting parameters (Eq.(2)). \textbf{(d)} $G_{\mathrm{S}}$ and $G_{\mathrm{N}}$ as a function of $V_\mathrm{G2}$ from (a) at $V_{\mathrm{SD2}}= -30\,\mathrm{\micro V}$ and $V_{\mathrm{SD2}}= -350\,\mathrm{\micro V}$, respectively.} 
	\label{fig:NIS_supplement}
\end{figure*}

	In order to compare the differential resistance of nanowire A with the theoretically predicted conductance in an Andreev enhanced QPC (Eq. 2) we repeated the measurement presented in Fig.~\ref{fig:NIS}a with a DC-setup. The data is smoothed over 30 steps and the differential conductance is calculated numerically. In addition the differential resistance and the voltage drop across the device is corrected for a constant line resistance of $R_{\mathrm{line}} = \,4.9\,\mathrm{k\Omega}$ (Fig.~\ref{fig:NIS_supplement}). The in-gap conductance $G_{\mathrm{S}}$ and normal conductance $G_{\mathrm{N}}$ are calculated as the average of $G(V_{\mathrm{SD2}})$ in the range $-60\,\mathrm{\micro V} < V_{\mathrm{SD2}} < 60\,\mathrm{\micro V}$ and  $-350\,\mathrm{\micro V} < V_{\mathrm{SD2}} < -250\,\mathrm{\micro V}$, respectively.
	
\subsection{Temperature dependent IV-curves}

	IV-curves at fixed gate voltage $V_{\mathrm{G3}} = 1.5\,\text{V}$ were measured and as function of temperature for nanowire C. The data were used to extract values for Fig.~\ref{fig:SIS}e.
	
\begin{figure}[h!]
	\includegraphics{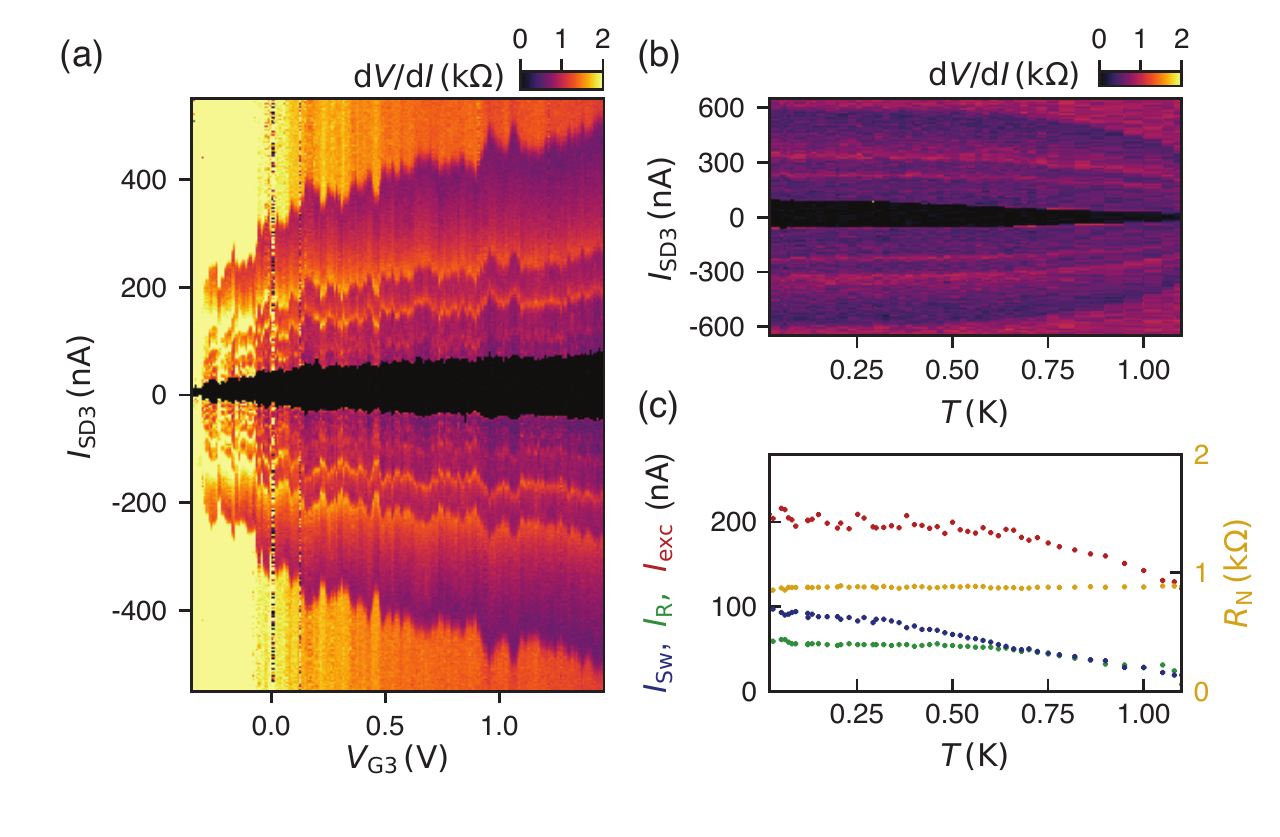}
	\caption{\label{fig:JJ_supplement}. \textbf{(a)} Differential resistance $\mathrm{d}V/\mathrm{d}I$ as a function of applied current $I_{\mathrm{SD3}}$ and gate voltage $V_{\mathrm{G3}}$ at $T=20\,\mathrm{mK}$ for nanowire C. The dataset is used to extract the junction parameters in Fig.~\ref{fig:SIS}c. \textbf{(b)}  $\mathrm{d}V/\mathrm{d}I$ as a function of applied current $I_{\mathrm{SD}}$ and sample temperature $T$ for nanowire C at $V_{\mathrm{G3}} = 1.5\,\text{V}$.  Switching current $I_{\mathrm{Sw}}$, retrapping current $I_{\mathrm{R}}$, excess current $I_{\mathrm{exc}}$ and normal state resistance $R_{\mathrm{N}}$ extracted from \textbf{(b)}.}
\end{figure}
	
\subsection{Josephson junction characteristics}

	The junction characteristics for 6 devices were measured and a summary of extracted values for $V_{\mathrm{G3}} = 1.5\,\mathrm{V}$ can be found in table~\ref{tab:jj-table}. Here, data in the main part of the paper were taken on nanowire A (Fig.~\ref{fig:NIS}), nanowire B (Fig.~\ref{fig:SIS}a-b) and nanowire C (Fig.~\ref{fig:SIS}c-d and Fig.~\ref{fig:MAR}). All measured devices display similar junction parameters with the exception of device A that shows a non-hysteretic I-V characteristic and device F that has a significantly lower $I_{\mathrm{Sw}}R_{\mathrm{N}}$ product. The $Z-\mathrm{parameter}$ and junction transparency $\mathcal{T}$ were extracted with the normalized excess current e$I_{\mathrm{exc}}R_{\mathrm{N}}/\Delta$ following Ref.~\cite{Flensberg1988}.
	
	\begin{table*}[h!]
		\caption{
			Characteristic Josephson junction parameters extracted for all measured devices at a gate voltage $V_{\mathrm{G3}} = 1.5\,\mathrm{V}$.}
		\begin{ruledtabular}
			\begin{tabular}{cccccccccc}
				nanowire & orientation & $w\,\text{(nm)}$  & $I_{\mathrm{Sw}}\,(\text{nA})$ & $I_{\mathrm{Sw}}R_{\mathrm{N}}\, (\mathrm{\micro V)}$ & e$I_{\mathrm{exc}}R_{\mathrm{N}}/ \Delta$ & $I_{\mathrm{R}}$/$I_{\mathrm{Sw}}$  & $Z$ & $\mathcal{T}$ \\ 
				\hline 
				A & [100] & 245 & 123.5 & 84.6 & 0.69 & 0.98 & 0.71 & 0.67  \\  
				\hline 
				B & [110] & 325  & 80.7 & 92.1 & 1.01 & 0.56 & 0.57 & 0.75 \\ 
				\hline 	
				C & [100] & 210 & 94.5 & 91.5 & 1.05 & 0.65 & 0.55 & 0.76  \\ 
				\hline 		
				D& [110] & 240  & 36.0 & 71.8 & 0.91 & 0.47 & 0.61 & 0.73  \\ 
				\hline 
				E & [110] & 210  & 61.7 & 87.2 & 1.18 & 0.56 & 0.51 & 0.80 \\  
				\hline 
				F & [100] & 310  & 17.0 & 47.5 & 0.59 & 0.51 & 0.75 & 0.64\\ 
				
			\end{tabular}
		\end{ruledtabular}
		\label{tab:jj-table}
	\end{table*}
	
\subsection{Multiple Andreev reflection data}

	In order to analyze the features in the subgap region of the dataset shown in Fig.~\ref{fig:SIS}c, the current-bias axis was mapped to the voltage drop across the junction $V_{\mathrm{SD3}}$ (cf. Fig.~\ref{fig:SIS}a). As expected for multiple Andreev reflection features the peaks in differential resistance are independent of the gate voltage $V_{\mathrm{G3}}$ and appear at voltages $V_{\mathrm{nm}}$ with $eV_{\mathrm{nm}} = 2\Delta/m$, where $m$ denotes the MAR order (Fig.~\ref{fig:MAR}b)
	
\begin{figure}[h]
	\includegraphics{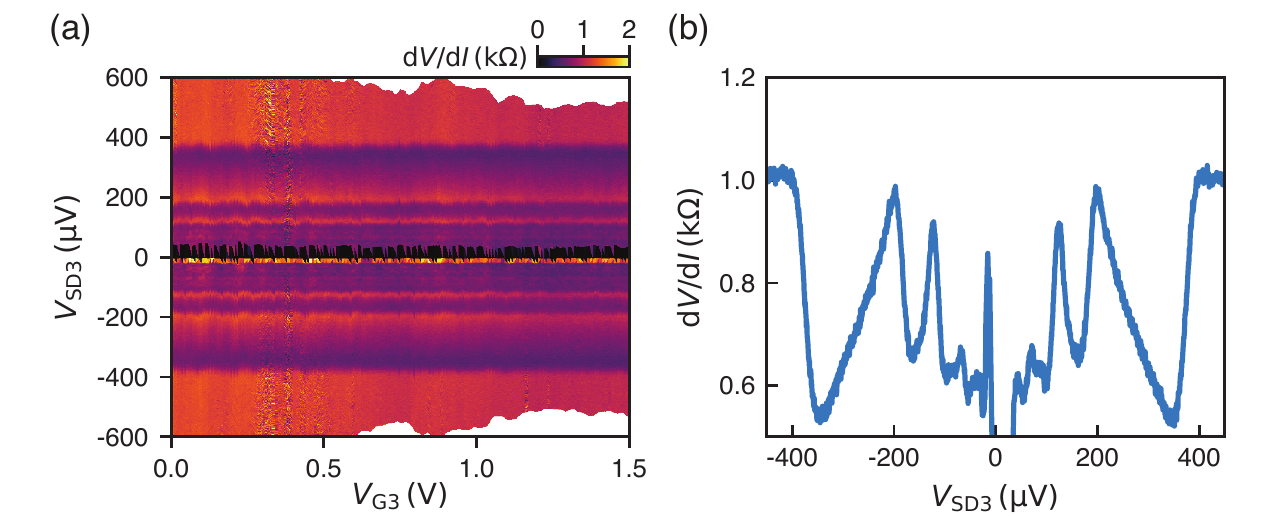}
	\caption{\label{fig:Fig5} \textbf{(a)} Differential resistance $\mathrm{d}V/\mathrm{d}I$ of the Josephson junction of nanowire C as a function of voltage drop across the junction $V_{\mathrm{SD3}}$ and gate voltage  $V_{\mathrm{G3}}$. \textbf{(b)} Averaged $\mathrm{d}V/\mathrm{d}I$ in the range $1.3\,\text{V} < V_{\mathrm{G3}} < 1.5\,\text{V}$. Peaks in the differential resistance are identified as signatures of multiple Andreev reflections.}
\end{figure}

\end{document}